\documentclass[11pt]{article}

\usepackage{PRIMEarxiv}

\usepackage[utf8]{inputenc} % allow utf-8 input
\usepackage[T1]{fontenc}    % use 8-bit T1 fonts
\usepackage{hyperref}       % hyperlinks
\usepackage{url}            % simple URL typesetting
\usepackage{booktabs}       % professional-quality tables
\usepackage{amsfonts}       % blackboard math symbols
\usepackage{nicefrac}       % compact symbols for 1/2, etc.
\usepackage{microtype}      % microtypography
\usepackage{lipsum}
\usepackage{fancyhdr}       % header
\usepackage{graphicx}       % graphics
\graphicspath{{media/}}     % organize your images and other figures under media/ folder
\usepackage{booktabs}
\usepackage{adjustbox}
\usepackage{comment}
\usepackage{bm}
\usepackage{multirow}
\usepackage{threeparttable}
\usepackage{moreverb}
\usepackage{amsmath}

\title{A Bayesian likely responder approach\\
for the analysis of randomized controlled trials
}

\author{
  Annan Deng \\
  Department of Population Health \\
  NYU School of Medicine \\ 
  New York, NY, USA\\
  \texttt{ad6557@nyu.edu} \\
  \And
  Carole Siegel \\
  Department of Psychiatry\\ 
  NYU School of Medicine\\ 
  New York, NY, USA\\
  \texttt{Carole.Siegel@nyulangone.org}
  \And
  Hyung G. Park \\
  Department of Population Health \\
  NYU School of Medicine \\
  New York, NY, USA\\
  \texttt{parkh15@nyu.edu} \\
}

\begin{document}
\maketitle

\begin{abstract}
An important goal of precision medicine is to personalize medical treatment by identifying individuals who are most likely to benefit from a specific treatment. The Likely Responder (LR) framework, which identifies a subpopulation where treatment response is expected to exceed a certain clinical threshold, plays a role in this effort. However, the LR framework, and more generally, data-driven subgroup analyses, often fail to account for uncertainty in the estimation of model-based data-driven subgrouping. We propose a simple two-stage approach that integrates subgroup identification with subsequent subgroup-specific inference on treatment effects. We incorporate model estimation uncertainty from the first stage into subgroup-specific treatment effect estimation in the second stage, by utilizing Bayesian posterior distributions from the first stage. We evaluate our method through simulations, demonstrating that the proposed Bayesian two-stage model produces better calibrated confidence intervals than naïve approaches. We apply our method to an international COVID-19 treatment trial, which shows substantial variation in treatment effects across data-driven subgroups.
\end{abstract}


\section{Introduction}

Precision medicine aims to tailor medical treatment to the individual characteristics of each patient \cite{national2011toward}.
Tracking individual differences in treatment effects has become increasingly important, as many treatments do not uniformly benefit all patients \cite{kaiser2022heterogeneity,sundstrom2023heterogeneity}.
In randomized controlled clinical trials (RCTs), there are instances where no significant differences in efficacy are observed between candidate therapies and a placebo when evaluating the group as a whole. However, this can obscure individual variations, with some participants responding well, while others derive little benefit, leading to a minimal apparent overall treatment effect \cite{ruberg2010mean}.

Laska et al.\cite{laska2022likely} introduced  
the \emph{Likely Responder} (LR) framework to identify patients predicted, from pre-treatment characteristics,  to achieve a favorable outcome under a given treatment and to conduct inference on treatment effects within this predicted LR subgroup. By first identifying these likely responders and then comparing treatment versus control within this subgroup, the LR framework seeks to provide a clinically interpretable, targeted assessment of benefit   \cite{laska2022likely,laska2020gabapentin,wastyk2023randomized,kalso2007predicting}. 
Our work builds on this line of research by addressing a key gap: propagating uncertainty from subgroup identification into subsequent subgroup-specific causal inference.
% \st{By leveraging individual-specific features associated with treatment response, the LR framework aligns with the goals of personalized medicine—optimizing treatment assignment based on patient characteristics.}   

\begin{figure}[bt]
\centering
\includegraphics[width=14cm]{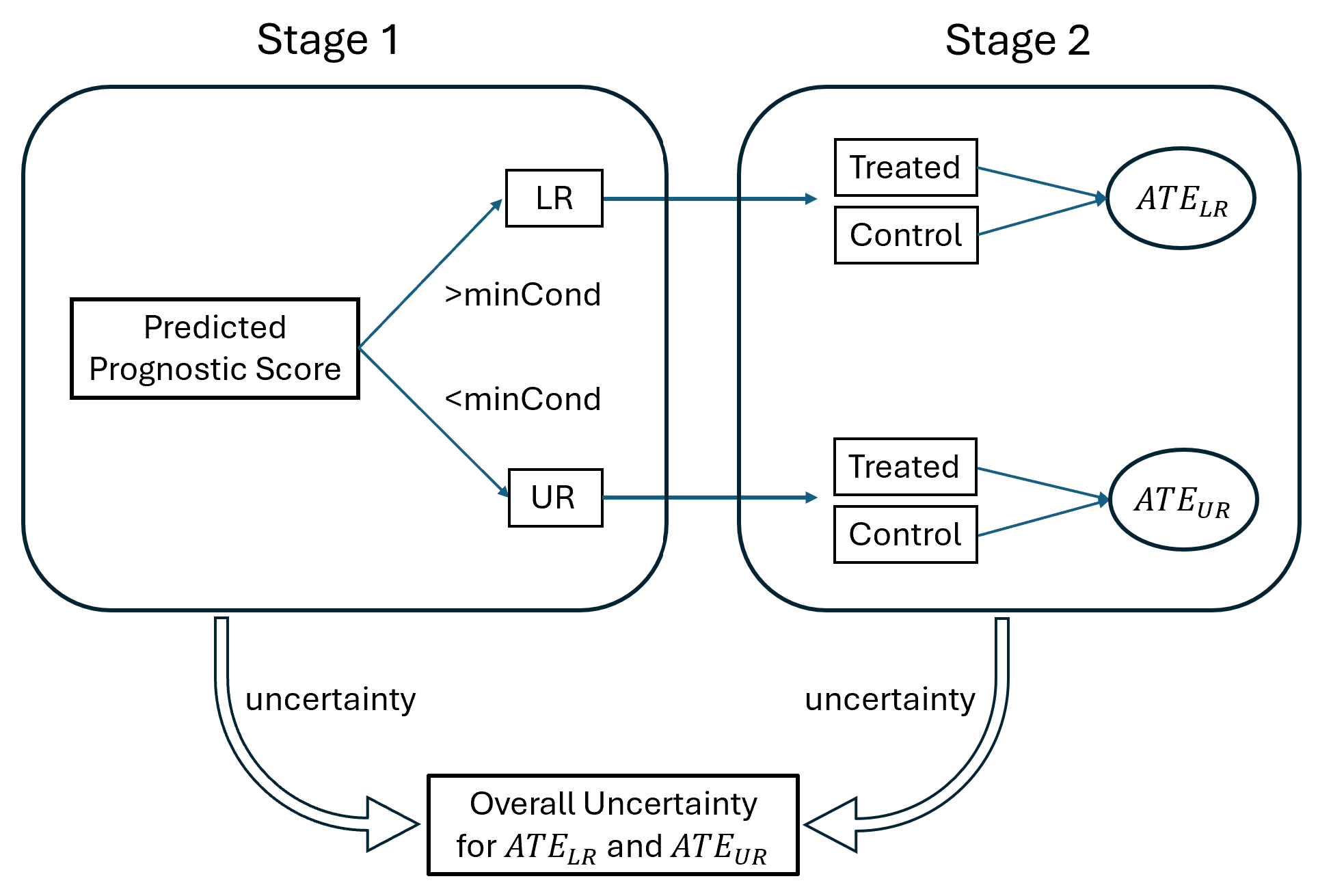}
\caption{Diagram of a two-stage likely responder framework design. In Stage 1, LR and UR represent likely responder and unlikely responder subgroups, respectively. In Stage 2, $\text{ATE}_{LR}$ and $\text{ATE}_{UR}$ represent the Average Treatment Effect in the LR and UR subgroups. \texttt{minCond} is set pre-trial, corresponding to a clinically meaningful threshold to define LR vs. UR.}
\label{diagram}
\end{figure}

Likely responders are identified using a prognostic score \cite{hansen2008prognostic}, which models the outcome as a function of baseline covariates within a treatment group.
This score can be viewed as a \emph{prognostic balancing score} (PBS), as individuals with similar PBS values are approximately ``balanced’’ on their covariates;  consequently, PBS values can be imputed counterfactually for patients in the comparator arm from their baseline characteristics. 
Patients are classified as likely responders when their PBS exceeds a prespecified threshold, \texttt{minCond}, chosen to reflect a clinically meaningful outcome level \cite{laska2022likely,laska2020gabapentin}. 
Different clinical priorities may motivate different \texttt{minCond} values (e.g., stricter thresholds for high-risk therapies), with each choice yielding a distinct LR analysis. When the PBS model is adequate, inference proceeds by comparing outcomes between treatment and comparator \emph{within} the subgroup defined by the thresholded PBS.
Accordingly, treatment effect estimation under the LR framework follows a two-stage workflow: (i) construct the PBS and apply the clinical threshold to define LR membership, and (ii) estimate the subgroup-specific treatment effect within the LR set. A diagram illustrating this workflow is shown in Fig. \ref{diagram}.

In Fig. \ref{diagram}, the uncertainty from both model-based prognostic score estimation in Stage 1 and treatment effect estimation in Stage 2 contribute together to the overall uncertainty of subgroup treatment effect uncertainty estimation. Both sources contribute to the sampling variability of the subgroup-specific effect estimate.
However, previous applications that  estimate the average treatment effect (ATE) among LRs typically \emph{ignore} uncertainty from the  prognostic-score estimation in subsequent downstream analyses for evaluating the subgroup-specific treatment effects \cite{laska2020gabapentin,laska2022likely,chalkou2021two,goligher2023heterogeneous}. This oversight can lead to potentially overconfident inference on the subgroup-specific treatment effects. 
The need to propagate first-stage uncertainty is recognized in adjacent literatures. For example, in observational studies, investigators have used Bayesian frameworks to carry forward uncertainty from estimated propensity scores into outcome analyses  \cite{liao2020uncertainty,alvarez2021uncertain}. Our contribution adapts this uncertainty-propagation principle to a different setting: quantifying subgroup-specific treatment effects in a \emph{predictive} framework, where prognostic scores are used to define clinically interpretable LR subgroups rather than to adjust for treatment-confounding. 

We propose an approach to enable the propagation of uncertainty that consists of two stages. 
Firstly, in the ``design’’ stage (Stage 1), we draw from the posterior of the prognostic score model and, for each draw, induce a subgrouping by applying the prespecified threshold to the sampled PBS scores. Each posterior draw thus defines a \emph{posterior-induced design}. Subsequently, in the ``evaluation’’ stage (Stage 2), for each design, we estimate the subgroup-specific ATE under each realization of the prognostic score-based subgrouping implementation. We then pool the resulting subgroup-specific ATEs across posterior-induced designs, thereby quantifying the overall uncertainty by accounting for the uncertainty in both (subgroup identification and treatment effect computation) stages of statistical evaluation.

To validate our proposed methodology, we conduct simulations with nonlinear predictor-outcome relationships and both continuous and binary outcomes. We compare results between a conventional approach, which uses point estimates in the design stage followed by a generalized linear model (GLM) based on the single stratification, and our two-stage Bayesian approach, which draws posterior samples in the design stage and estimates the ATE via GLM for each posterior-induced stratification in the evaluation stage.
Reproducible code for the two-stage Bayesian procedure is available at \url{https://github.com/ad6557/Two-stage-Bayesian-ATE}. 
We illustrate the utility of this framework through its application to the COMPILE study (Continuous Monitoring of Pooled International Trials of Convalescent Plasma for COVID-19 Hospitalized Patients) \cite{troxel2022association}. By explicitly accounting for posterior uncertainties in data-driven subgrouping, our methodology aims to improve the assessment of subgroup-specific treatment effects and support more informed medical interventions.

\section{Two stage Model}

\subsection{Estimands and Assumptions} \label{2.1}

We consider an RCT with $n$ independent participants, randomly assigned to treatment ($T=1$) or to control ($T=0$). For each individual $i=1,\dots,n$, we observe ($Y_i$, $X_i$, $T_i$), where $Y_i$ is the outcome and $X_i\in\mathbb{R}^p$ are baseline covariates. Let $\bm{Y}=(Y_1,\ldots,Y_n)^\top$ and $\bm{T}=(T_1,\ldots,T_n)^\top$, and $\bm{X}=(X_1^\top,\ldots,X_n^\top)^\top$.  
Each outcome $Y$ is assumed to follow a distribution from the exponential family given $T$ and $X$. The conditional mean $E(Y \mid T, X)$ is modeled as:
\begin{equation}
h(E(Y \mid T, X)) = f(X,T), 
\label{eq1}
\end{equation}
where $f(\cdot)$ is an unknown function that we infer; and $h(\cdot)$ is the canonical link function in the framework of GLMs, for example, the identity function for a Gaussian model and the logit function for a binomial model.

\paragraph{Assumptions.}
We assume: (i) Consistency: 
$E(Y_i \mid X_i, T_i = t) = E(Y_i^{(t)} \mid X_i, T_i = t), \ \text{for } i = 1, 2, \ldots, n \text{ and } t = 0, 1$; (ii)  Ignorability: 
$(Y_i^{(0)},Y_i^{(1)}) \perp T_i \mid X_i$; (iii) Positivity: 
$0<P(T_i=1 \mid X_i=x_i)<1$ (on the support of $X$); (iv) Stable unit treatment value assumption (SUTVA):  
$Y_i = Y_i^{(1)} I(T_i = 1)+Y_i^{(0)} I(T_i = 0)$. 

 Assumption (i) ensures that the potential outcome ($Y_i^{(t)}$) under the observed exposure $(T=t)$ aligns with the observed outcome $(Y_i)$, allowing us to use observed data to infer potential outcomes in the estimand calculation.
Assumption (ii), the no unmeasured confounders assumption, is naturally satisfied in an RCT.
Assumption (iii) requires excluding individuals who would always or never receive treatment. 
Finally, Assumption (iv) asserts that an individual’s outcome is unaffected by the treatment assignment of others (i.e., no interference).

\paragraph{Prognostic balancing score (PBS) and subgrouping.}
For the active treatment ($T{=}1$), the prognostic balancing score for individual $i$ is
\[
s_i \;=\; \mathbb{E}\!\left\{Y_i \mid T_i{=}1,\,X_i\right\},\qquad i=1,\ldots,n,
\]
i.e., the expected outcome under treatment given baseline covariates $X_i$. Let
$\mathbf{s}=(s_1,\ldots,s_n)^\top \in \mathbb{R}^n$ denote the vector of PBS values.
Patients are partitioned into $S$ strata (e.g., $S{=}2$ for LR vs.\ UR) by thresholding
$\mathbf{s}$ at prespecified clinical cutpoints. We write
\[
\boldsymbol{\nu}=(\nu_1,\ldots,\nu_n)^\top,\qquad \nu_i \in \{1,\ldots,S\},
\]
for the resulting subgroup labels, where $\nu_i$ indexes the stratum to which subject $i$ belongs.

We denote the causal treatment effect within a subpopulation $\psi \subseteq \Omega$ using the symbol $\Delta(\psi)$, which we define as: 

\begin{equation}
    \Delta(\psi)  = E[h(E(Y_i^{(1)} \mid X_i)) - h(E(Y_i^{(0)} \mid X_i))| i \in \psi],
\label{eq.delta}
\end{equation}
where the set $\psi$ can represent, for example, a likely responder (LR) or unlikely responder (UR) subgroup.

\subsection{Design Stage} \label{2.2}

While other Bayesian approach could be used, we employ Bayesian Additive Regression Trees (BART) to estimate $s_i=E\left(Y_i \mid T_i=1, X_i\right)$ in the ``design’’ stage \cite{chipman2010bart}. BART is a semi-parametric, ensemble-of-trees model that flexibly captures nonlinearity and high-order interactions via a sum of shallow regression trees with shrinkage priors. This structure accommodates many covariates and complex effect shapes while regularization mitigates overfitting. The Bayesian formulation yields posterior draws of the prognostic score, $\{s_i^{(k)}\}_{k=1}^K$, which we later use to induce posterior-driven subgroup designs. Implementation details and prior specifications are provided in Section A.2 of the Supplementary Material. %(See supporting information in in Section A2 in the Supplemental Material). Embedding this algorithm within a Bayesian inferential framework allows for uncertainty quantification, and helps prevent overfitting by ensuring that each tree contributes only a small part to the overall fit.

We approximate the regression function $f(\cdot)$ in Eq.~\ref{eq1} for $T=1$ with a sum-of-trees model to estimate PBS: 
$$
E(Y \mid T=1, x) =h^{-1}[f(x, T=1)]= h^{-1}(\sum^m_{j=1} g_j(x ; \mathrm{Tree}_j, M_j)),
$$
where, 
among $m$ trees in total, each $\mathrm{Tree}_j$ denotes a binary tree consisting of a set of interior node decision rules and a set of terminal nodes, and $M_j$ denotes a set of parameter values associated with $\mathrm{Tree}_j$.

Although estimated based on the data obtained from individuals with $T=1$,
the developed model for PBS can be applied to all individuals, including those with $T=0$, to evaluate PBS for all subjects:
$$
s_i = h^{-1}[ f(x_i, T=1) ]= h^{-1}[ \sum^m_{j=1} g_{ij}(x_i ; \mathrm{Tree}_{ij}, M_{ij}) ],\quad i=1,2, \ldots, n, 
$$
in which the uncertainty in the estimation of 
$f(\cdot, T=1)$ in this ``design’’ stage is incorporated through the posterior distribution of $s_i$.   
Given a prespecified subgrouping rule, we set:
$$\nu(X)  = \mbox{subgrouping} (h^{-1}[f(X, T=1)]),$$
where the subgrouping function $\nu(\cdot)$, defined through the operation ``subgrouping", deterministically maps pre-treatment characteristics $X \in \mathbb{R}^p$ to one of the predicted treatment response strata $\{1,\ldots,S\}$, given the estimated regression function $f(\cdot, T=1)$.

\subsection{Evaluation Stage}
Let us denote the observed data collectively as $\mathcal{D} = \{(X_i,Y_i,T_i), i=1,2, \ldots, n\}$. 

The causal estimand, $\Delta(\nu_0)$, of interest depends on the ``true" stratification denoted as $\nu_0$. Under the standard causal inference assumptions (i)-(iv), the casual estimate is written as:

%\begin{equation}
%    \Delta(\nu_0) = h[E(Y \mid T=1, \nu_0)] - h[E(Y \mid T=0, \nu_0)]
%\label{eq1.nu0xx}
%\end{equation}

\begin{equation}
    \Delta(\nu_0) = E[h\left(E(Y \mid T=1, X)\right) - h\left(E(Y \mid T=0, X)\right)| \nu_0]
\label{eq1.nu0}
\end{equation}

However,  the ``true" $\nu_0$ is unknown: the subgroup labels are induced by the Stage 1 prognostic model and the prespecified threshold. Let $s^{(k)}(\cdot)$ denote the $k$th posterior draw of the prognostic score and let
$$
\nu^{(k)} \;=\; \nu\big(s^{(k)}(\cdot)\big)
$$
be the corresponding \emph{posterior-induced design}  (deterministic given $s^{(k)}$ and the threshold).  This induces a posterior distribution over designs, $p\left(\nu\mid\mathcal D,\nu_0\right)$, via the Stage 1 posterior. 
%However,  the ``true" $\nu_0$ is unknown, thus we conduct Bayesian inference on $\nu_0$ in the design stage, and quantify uncertainty.Let us denote each posterior realization of $\nu_0$ as $\nu$, and the corresponding posterior distribution $p(\nu | \mathcal{D},\nu_0)$ given the data $\mathcal{D}$, will be utilized in estimating  $\Delta(\nu_0)$ in Eq.(\ref{eq1.nu0}). 
Specifically, 
in the ``evaluation’’ stage, we evaluate $\Delta(\nu_0)$ in Eq.~\ref{eq1.nu0} based on: 
\begin{equation}
    \Delta(\nu_0) = \int_{\nu} \Delta(\nu) p(\nu | \mathcal{D},\nu_0) d\nu
\label{eq2.nu0}
\end{equation}
where the integral is approximated based on  posterior samples drawn from $p(\nu | \mathcal{D},\nu_0)$.

\section{Computational procedure for Two-Stage Sampling}
In this section, we describe the proposed  two-stage sampling method to infer on $\Delta (\nu_0)$ in Eq.~\ref{eq1.nu0}.

\subsection{Drawing from the posterior predictive distribution of the ``design’’ \texorpdfstring{$\nu$}{nu} in the first stage}\label{sec3.1}

In the first (``design’’) stage, multiple ($K$) draws are taken from posterior distribution of the balancing score, $p(s\mid \mathcal{D}, \nu_0)$, yielding the corresponding $K$ draws from the posterior distribution of the ``design’’ $p(\nu\mid \mathcal{D})$, using the following procedure.

\begin{enumerate}
    \item We partition the $n$ subjects into two sets: 1) ``design’’; and 2) ``evaluation’’ sets. The ``design’’ set can comprise one half of the individuals with $T=1$ (let us denote the corresponding sample size as $n_{\text{design}}$), and the ``evaluation’’ set comprises the remaining individuals with $T=1$ and those with $T=0$ (let us denote the corresponding sample size as $n_{\text{eval}}=n - n_{\text{design}}$). 
     
    \item For the proposed two-stage approach, we employ a Bayesian model (e.g., BART) to train the PBS model on the $n_{\text{design}}$ treated subjects in the ``design’’ set, and then apply the trained predictive model on subjects in the ``evaluation’’ set to obtain a sample of $\mathrm{K}$ draws from the posterior distribution of the PBS, with each draw being noted as $\boldsymbol{s}_k = [s_{k,1},\ldots,s_{k,n_{\text{eval}}}]^\top \in \mathbb{R}^{n_{\text{eval}}}$, $k=1,2, \ldots, K$.
    
    \item Then we use the $K$ draws from the posterior distribution of PBS to perform $K$ implementations of subgrouping, each representing a subgroup design  $\boldsymbol{\nu}_k \in \mathbb{R}^{n_{\text{eval}}}, k=1,2, \ldots, K$.\\
    Note, the subgrouping $\boldsymbol{\nu}_k$ is based on the value of the realized PBS. For example, given each posterior-sampled $\boldsymbol{s}_k$, the individual $i$ will be assigned to likely responder (LR) subgroup if $s_{k,i}>minCond$, otherwise be assigned to unlikely responder (UR) subgroup.

\end{enumerate}

In contrast, a naïve model would train the PBS model on the $n_{\text{design}}$ treated subjects in the ``design’’ set, and then apply the trained model on subjects in the ``evaluation’’ set to get a single point estimate of the balancing score vector, denoted as $\hat{\boldsymbol{s}} \in \mathbb{R}^{n_{\text{eval}}}$, used to perform a single implementation of subgrouping. % denoted as $\nu$.

In summary, the output of the first stage (the design module), produces either $K$ draws (for the proposed two-stage method) from posterior distribution $p(\nu\mid \mathcal{D})$, each representing a different design, or a single point estimate (for ``naïve’’ method).

\subsection{Drawing from the conditional posterior distribution of $\Delta(\nu_k)$ and combining across the designs $\nu_k$ $(k=1,\ldots,K)$ in the second stage}

Once the PBS-based design $\nu_k$ categorizes each individual $i$ into one of the $S$ subgroups $\nu_k$, in the evaluation stage, we employ GLM to estimate the ATE for each subgroup, using the subgroup-specific mean models of the form: $h(E(Y_i | T_i)) = \beta_{0\nu_k} + \beta_{\nu_k} T_i$, where subjects $i$ are stratified by subgrouping $\nu_k$ for each $k$. Using this GLM in the evaluation stage, we can obtain MLE of the causal estimand in Eq.~\ref{eq1.nu0} at $\nu_0 = \nu_k$, which we denote as $\hat{\Delta}_k(\nu_k)$ $(k=1,\ldots,K)$ and its variance which we denote as $\hat{\sigma}_k^2$ $(k=1,\ldots,K)$, which will be combined by Rubin's rule (see below).

On the other hand, in the naïve model, the design stage generates a single $\nu$, which leads to one MLE estimate for $\Delta$ in the evaluation stage.

\subsection{Between- and within-design uncertainty}

Viewing the subgrouping $\nu$ as a missing value that can be multiply-imputed, we estimate the posterior variance of $\Delta(\nu_0)$ using Rubin's combining rules \cite{rubin1987multiple}.

\begin{equation}
    \begin{aligned}
    \operatorname{Var}(\Delta(\nu_0) \mid \mathcal{D}) 
    &= \mathbb{E}(\operatorname{Var}(\hat{\Delta}(\nu_0) \mid  \nu_0 = \nu, \mathcal{D}) \mid \mathcal{D}) + \operatorname{Var}(\mathbb{E}(\hat{\Delta}(\nu_0) \mid  \nu_0 = \nu, \mathcal{D}) \mid \mathcal{D}) \\
    &\approx \bar{\sigma}_K^2 + \left(1+\frac{1}{K}\right) B_K^2 \\
    &= \frac{\sum_{k=1}^K \sigma_k^2}{K} + \left(1+\frac{1}{K}\right) \frac{\sum_{k=1}^K\left(\Delta_k(\nu_k)-\bar{\Delta}_K\right)^2}{K-1}, 
    \end{aligned}
\label{rubin}
\end{equation}
where $\bar{\sigma}_K^2  = \frac{\sum_{k=1}^K \hat{\sigma}_k^2}{K}$ 
and 
$B_K^2 =  \frac{\sum_{k=1}^K\left(\hat{\Delta}_k(\nu_k)-\bar{\Delta}_K\right)^2}{K-1}$. 

Here $\hat{\Delta}_k(\nu_k)$ means the estimated causal estimand given the subgrouping $\nu_k$ and 
$\bar{\Delta}_K$ means the average causal estimand across all $K$ stratifications, where we approximate the distribution of $\nu$ using the posterior samples $\nu_1,\ldots, \nu_K$. 

The ``within-design" variability, $\bar{\sigma}_K^2$, quantifies the variability of $\Delta(\nu_0)$. The second term $(1+\frac{1}{K}) B_K^2$ corresponds to the ``Between-design" variability across the samples from the posterior distribution of $\nu$. Naïve method which only gets one point estimate from the ``design’’ stage ignores the between-design variability, will leads to the underestimation of overall variability.

\section{Simulations}\label{sec3}

\subsection{Data-generating mechanisms}\label{subsec1}

In our simulation illustration, we considered two outcome types: continuous and binary. For each outcome scenario, we generated a 10-dimensional covariate vector $X_i$ independently for each individual under two settings. In the first setting, all covariates were drawn from a multivariate Gaussian distribution with mean 0 and variance 1. In the second setting, $X_i$ consisted of a mixture of variable types: 4 Gaussian, 2 binary (Bernoulli with success probability 0.5), 2 exponential, and 2 chi-square ($df=10$). Treatment indicators $T \in \{0,1\}$ were assigned independently from a Bernoulli(0.5) distribution. 
We generated the outcomes based on the canonical parameters $f(\boldsymbol{x}_i, T_i)$ in Eq.~\ref{eq1} by setting:
$f(\boldsymbol{x}_i, T_i) = \boldsymbol{\mu}^\top \boldsymbol{x}_i + g(\boldsymbol{x}_i) \cdot T_i,$
where $\boldsymbol{\mu} = [.5, 1, .75, 1, .5, -.5, -1, -.75, -1, -.5]$. 
For binary outcomes, following the Bernoulli distribution, we defined:
$g(\boldsymbol{x}_i) = \left( \frac{\sin(\pi x{i1})}{5} - \frac{\sin(\pi x_{i2})}{5} + \frac{x_{i3}^2}{5} - \frac{x_{i4}^2}{5} - \frac{x_{i5} x_{i6}}{10} + x_{i7} - x_{i8} + \frac{\exp(x_{i9})}{5} - \frac{\exp(x_{i10})}{5} \right).$
For continuous outcomes, we defined:
$g(\boldsymbol{x}_i) = \left( \frac{\sin(\pi x_{i1} x_{i2})}{2} + \frac{(x_{i3} - 0.5)^2}{3} + \frac{x_{i4}^3}{4} + \frac{x_{i5}}{5} - \frac{\sin(\pi x_{i6} x_{i7})}{2} - \frac{(x_{i8} - 0.5)^2}{3} - \frac{x_{i9}^3}{4} - \frac{x_{i10}}{5} \right)$
and the additive Gaussian noise to $f(\boldsymbol{x}_i, T_i)$ conforms to $\epsilon_i \sim N(0, 1)$. 

We conducted simulation for various sample sizes, $n \in \{500, 1000, 2000\}$, corresponding to design stage sample sizes $n_{\text{design}}$ of $\{125, 250, 500\}$ and evaluation stage sample size $n_{\text{eval}}$ of $\{375, 750, 1000\}$, as the model is trained on half of the subjects with $T_i=1$ during the ``design’’ stage and the remaining samples $n_{\text{eval}}$ with $T_i=0$ or $1$ are used in the evaluation stage.
For each $n$, we simulated $R=200$ datasets. 

For each outcome scenario, true values of the marginal estimand, $\Delta$ in Eq.~\ref{eq1.nu0}, are determined by simulating a set of $10^7$ subjects, using the same covariate and outcome distributions in the simulation study. Then, using the same covariates and assuming that all subjects receive treatment, we calculate $Y^{(1)}$ for each individual. Subjects are classified into the LR or UR groups based on whether their $Y^{(1)}$ values exceed or fall below the predefined threshold, \texttt{minCond}. The estimand is then computed by calculating the difference in the average outcomes between the treated and untreated subjects, following Eq.~\ref{eq.delta}.
For a binary treatment response $Y_i \in \{0,1\}$, we define $\Delta$ in terms of the log odds ratio (log OR) using the using the logit link function $h$, and
the \texttt{minCond} threshold was set at 0.5 at probability scale; 
For a continuous treatment response $Y_i \in \mathbb{R}$, the link function $h$ is identity function, and
we set \texttt{minCond} to 0.

\subsection{Competing Methods}

For the ``design’’ stage that identifies LR and UR subgroups, we consider two approaches:  eXtreme Gradient Boosting (XGBoost) \cite{chen2016xgboost} and BART \cite{chipman2010bart}. 
XGBoost is an optimized implementation of Gradient Boosting Machines (GBMs), a widely used ensemble learning technique \cite{friedman2001greedy}.
GBMs iteratively refine predictions by combining weak learners through gradient descent. XGBoost enhances this process with regularization, parallelized computation, efficient handling of missing values, and flexible loss functions, while further improving performance through tree depth control, subsampling, and feature selection.
The details of GBM is provided in Section A1 in the Supplemental Material.

Our two-stage sampling-based method is benchmarked against a ``naïve’’ approach, where either XGBoost or BART is employed as the first-stage model, but no correction is applied in the second stage to account for estimation uncertainty from the first stage. 
%\st{To demonstrate that the observed differences between the frequentist ``naïve’’ method and the two-stage method are not solely due to the choice of XGBoost versus BART in the first stage, we introduced a two stage ``naïve’’ method using BART. } 
The details of all methods are as follows:

\begin{enumerate} 
\item ``Naïve’’ method with XGBoost: Utilize XGBoost in the design stage to obtain a point estimate of PBS, $\hat{s}$ (see step 2 in Section \ref{sec3.1}), thus obtain one single implementation $\nu$ (see step 3 in Section \ref{sec3.1}). Then, apply GLM on this subgrouping implementation to compute the subgroup-specific treatment effect $\hat{\Delta}(\nu)$ and its uncertainty interval.

\item ``Naïve’’ method with BART: The same as the above ``naïve’’ method with XGBoost, but utilize the posterior mean of BART model as the point estimate $\hat{s}$.
  
\item Two stage Bayesian method (``Corrected"):  Calculate the marginal posterior mean of $\Delta$ in Eq.~\ref{eq.delta} for each subgroup the same as the above method with BART, however, the marginal posterior variance is corrected using Eq.~\ref{rubin}.

\end{enumerate}

The XGBoost models in the design stage were trained using the `xgboost' package in R \cite{chen2019package}, and BART models were trained using the `BART' package in R \cite{sparapani2021nonparametric}. 
%In the evaluation stage, we used 'beanz' package for inference of treatment effects, as it allows estimation of subgroup-specific treatment effects and their standard errors for all groups within a single function call \cite{wang2018beanz}. 
The simulation environment was coded in R v. 4.2.1. 
For continuous outcomes, XGBoost models were configured to minimize the squared error loss function with root mean squared error (RMSE) as the evaluation metric, while for binary outcomes, models employed logistic regression with the deviance
%log loss $- [ y \log(\hat{y}) + (1 - y) \log(1 - \hat{y}) ]$ 
as the evaluation metric. Hyperparameter tuning was performed via 5-fold cross-validation over a predefined grid to mitigate overfitting, adjusting learning rate ({0.02, 0.05, 0.1, 0.3}), maximum tree depth ({3, 5, 7}), and subsample ratio of features ({0.8, 1.0}). 
BART models were fitted using default hyperparameters, including a sum of 200 trees, a base prior of 0.95, and a power prior of 2, chosen to penalize excessive branch growth (see Section A2 in the Supplemental Material). Each model was trained using a single chain MCMC procedure, consisting of 500 burn-in iterations followed by 100 retained iterations for analysis.

\subsection{Simulation Results}

The performance of the proposed methods was compared against naïve approaches using several metrics, including mean squared error (MSE), variance, mean absolute bias, mean standard error, and 95\% Confidence Interval (CI) coverage for the true casual estimand $\Delta$. For each simulation run, denoted by $r$ ($r = 1, \ldots, R$), where $R=200$, we computed the estimator $\hat{\Delta}_r$ and its corresponding standard error $se_r$ to estimate the population parameter $\Delta$. 
The metrics were computed as follows:

\begin{itemize}

    \item $\mbox{Bias}(\hat{\Delta}_r) = \frac{1}{R}\sum_{r=1}^R \hat{\Delta}_r - \Delta$
    \item $\mbox{Var}(\hat{\Delta}_r) = \frac{1}{R}\sum_{r=1}^R (\hat{\Delta}_r - \bar{\Delta})^2,$where $ \bar{\Delta} = \frac{1}{R}\sum_{r=1}^R \hat{\Delta}_r$
    \item $\mbox{MSE}(\hat{\Delta}_r) = \mbox{bias}(\hat{\Delta}_r)^2 + \mbox{var}(\hat{\Delta}_r)$
    \item $\overline{SE}(\hat{\Delta}_r)=\frac{1}{R}\sum_{r=1}^R se_r$
    \item $\mbox{Covrg}(\hat{\Delta}_r) = \frac{1}{R}\sum_{r=1}^R I (\Delta \in [\hat{\Delta}_r - 1.96 \times se_r,\hat{\Delta}_r + 1.96 \times se_r])$
\end{itemize}

Tables \ref{tab1} and \ref{tab2} present the detailed results for binary and continuous outcomes. 
Regarding subgroup-specific treatment effect estimation, the results indicate that the ``naïve’’ method with BART slightly outperforms the ``naïve’’ method with XGBoost in terms of bias, variance, and MSE. 
Both XGBoost-trained and BART-trained models exhibit bias toward the null in subgroup-specific treatment effect estimation, though the bias in BART is slightly smaller than that in XGBoost. Additionally, the variance of the BART-trained model across 200 simulation runs is lower, indicating greater stability. As a result, the MSE is smaller for BART than for XGBoost. 
However, the average standard error (SE) of the ATE for both the ``naïve’’ frequentist method and the native method with BART is smaller than that of the proposed Bayesian uncertainty-corrected method, and their 95\% Confidence Interval (CI) coverages fall below 95\%, indicating that the ``naïve’’ methods underestimate the uncertainty in their estimates.
In contrast, the two-stage Bayesian method maintains strong precision and accuracy while providing coverage that is close to the nominal 95\%, demonstrating its superior performance in both uncertainty estimation and overall reliability.

\begin{table*}[t]
\caption{Simulation Results on Estimand for Binary Outcomes in the Gaussian and Mixed-Type Covariates Settings}
\label{tab1}

\begin{adjustbox}{max width=\textwidth}
\begin{tabular}{|c c | ccccc | ccccc | cc |}
\toprule 
\multirow{3}{*}{\shortstack{\textbf{\(\text{n}_{\text{eval}}\)} \\ \((\text{n}_{\text{design}})\)}} 
 & & \multicolumn{5}{c|}{\textbf{XGBoost} \ PBS} & \multicolumn{7}{c|}{\textbf{BART} \ PBS} \\
\cmidrule[1pt](r){3-7}
\cmidrule[1pt](r){8-14}
& & & & & \multicolumn{2}{c|}{\textbf{Naïve}} & & & & \multicolumn{2}{c|}{\textbf{Naïve}} & \multicolumn{2}{c|}{\textbf{Corrected}} \\
\cmidrule(lr){6-7}
\cmidrule(lr){11-12}
\cmidrule(lr){13-14}
& & MSE & Var & Bias & $\overline{SE}$ & Covrg 
    & MSE & Var & Bias & $\overline{SE}$ & Covrg 
    & $\overline{SE}$ & Covrg \\
\midrule

% ==================== Gaussian ====================
\multicolumn{14}{|c|}{\textbf{Gaussian covariates}} \\
\midrule
\multirow{2}{*}{\shortstack{$375$ \\ $(125)$}} 
& UR & 0.171 & 0.131 & -0.200 & 0.354 & 0.885 
      & 0.195 & 0.155 & -0.199 & 0.371 & 0.870 
      & 0.439 & 0.940 \\
& LR & 0.185 & 0.148 &  0.193 & 0.359 & 0.890 
      & 0.240 & 0.212 &  0.167 & 0.382 & 0.855 
      & 0.442 & 0.915 \\
\cmidrule{1-14}
\multirow{2}{*}{\shortstack{$750$ \\ $(250)$}} 
& UR & 0.097 & 0.078 & -0.140 & 0.258 & 0.890 
      & 0.100 & 0.092 & -0.089 & 0.268 & 0.905 
      & 0.311 & 0.965 \\
& LR & 0.090 & 0.077 &  0.115 & 0.257 & 0.920 
      & 0.085 & 0.074 &  0.106 & 0.262 & 0.910 
      & 0.303 & 0.950 \\
\cmidrule{1-14}
\multirow{2}{*}{\shortstack{$1500$ \\ $(500)$}} 
& UR & 0.049 & 0.042 & -0.089 & 0.185 & 0.880 
      & 0.042 & 0.036 & -0.080 & 0.188 & 0.925 
      & 0.215 & 0.965 \\
& LR & 0.047 & 0.038 &  0.094 & 0.184 & 0.910 
      & 0.044 & 0.040 &  0.068 & 0.188 & 0.910 
      & 0.215 & 0.950 \\
\midrule

% ==================== Mixed-type ====================
\multicolumn{14}{|c|}{\textbf{Mixed-type covariates}} \\
\midrule
\multirow{2}{*}{\shortstack{$375$ \\ $(125)$}} 
& UR & 0.223 & 0.156 & -0.259 & 0.360 & 0.850 
      & 0.164 & 0.157 & -0.085 & 0.374 & 0.920 
      & 0.442 & 0.970 \\
& LR & 0.179 & 0.133 &  0.215 & 0.352 & 0.885 
      & 0.144 & 0.142 &  0.046 & 0.367 & 0.930 
      & 0.431 & 0.965 \\
\cmidrule{1-14}
\multirow{2}{*}{\shortstack{$750$ \\ $(250)$}} 
& UR & 0.096 & 0.064 & -0.178 & 0.258 & 0.905 
      & 0.079 & 0.074 & -0.067 & 0.263 & 0.940 
      & 0.303 & 0.975 \\
& LR & 0.115 & 0.076 &  0.197 & 0.254 & 0.815 
      & 0.076 & 0.072 &  0.061 & 0.263 & 0.965 
      & 0.303 & 0.980 \\
\cmidrule{1-14}
\multirow{2}{*}{\shortstack{$1500$ \\ $(500)$}} 
& UR & 0.059 & 0.039 & -0.142 & 0.185 & 0.835 
      & 0.040 & 0.039 & -0.041 & 0.191 & 0.925 
      & 0.218 & 0.950 \\
& LR & 0.060 & 0.038 &  0.145 & 0.185 & 0.860 
      & 0.046 & 0.039 &  0.085 & 0.187 & 0.910 
      & 0.215 & 0.940 \\
\bottomrule
\end{tabular}
\end{adjustbox}

\begin{flushleft}
\textit{Note.} UR = unlikely responders; LR = likely responders; 
$n$ = total sample size per simulation run; 
$n_{\text{design}}$ = sample size used to train the model in the design stage.
“XGBoost PBS” and “BART PBS”: XGBoost or BART was used to estimate PBS for subgrouping. 
True treatment effects: Gaussian, $\Delta(\mathrm{UR}) = -0.512$, $\Delta(\mathrm{LR}) = 0.524$;  
Mixed-type, $\Delta(\mathrm{UR}) = -0.652$, $\Delta(\mathrm{LR}) = 0.686$.  
\end{flushleft}
\end{table*}

\begin{table*}[t]
\caption{Simulation Results on Estimand for Continuous Outcomes in the Gaussian and Mixed-Type Covariates Settings}
\label{tab2}

\begin{adjustbox}{max width=\textwidth}
\begin{tabular}{|c c | ccccc | ccccc | cc |}
\toprule 
\multirow{3}{*}{\shortstack{\textbf{\(\text{n}_{\text{eval}}\)} \\ \((\text{n}_{\text{design}})\)}} 
 & & \multicolumn{5}{c|}{\textbf{XGBoost} \ PBS} & \multicolumn{7}{c|}{\textbf{BART} \ PBS} \\
\cmidrule[1pt](r){3-7}
\cmidrule[1pt](r){8-14}
& & & & & \multicolumn{2}{c|}{\textbf{Naïve}} & & & & \multicolumn{2}{c|}{\textbf{Naïve}} & \multicolumn{2}{c|}{\textbf{Corrected}} \\
\cmidrule(lr){6-7}
\cmidrule(lr){11-12}
\cmidrule(lr){13-14}
& & MSE & Var & Bias & $\overline{SE}$ & Covrg 
    & MSE & Var & Bias & $\overline{SE}$ & Covrg 
    & $\overline{SE}$ & Covrg \\
\midrule

% ==================== Gaussian ====================
\multicolumn{14}{|c|}{\textbf{Gaussian covariates}} \\
\midrule
\multirow{2}{*}{\shortstack{$375$ \\ $(125)$}} & UR
& 0.214 & 0.193 & -0.144 & 0.366 & 0.890 
& 0.139 & 0.126 & -0.117 & 0.347 & 0.945 
& 0.406 & 0.960 \\
& LR  
& 0.186 & 0.159 &  0.165 & 0.369 & 0.900 
& 0.155 & 0.135 &  0.142 & 0.348 & 0.915
& 0.407 & 0.955 \\
\cmidrule{1-14}
\multirow{2}{*}{\shortstack{$750$ \\ $(250)$}} & UR 
& 0.103 & 0.088 & -0.124 & 0.249 & 0.890 
& 0.081 & 0.073 & -0.093 & 0.242 & 0.910 
& 0.272 & 0.940 \\
& LR
& 0.082 & 0.072 &  0.098 & 0.252 & 0.910 
& 0.068 & 0.063 &  0.067 & 0.244 & 0.940 
& 0.275 & 0.960 \\
\cmidrule{1-14}
\multirow{2}{*}{\shortstack{$1500$ \\ $(500)$}} & UR 
& 0.044 & 0.039 & -0.075 & 0.173 & 0.890 
& 0.036 & 0.033 & -0.060 & 0.169 & 0.905 
& 0.184 & 0.930 \\
& LR
& 0.038 & 0.033 &  0.072 & 0.173 & 0.915 
& 0.034 & 0.030 &  0.068 & 0.169 & 0.905 
& 0.185 & 0.940 \\
\midrule

% ==================== Mixed-type ====================
\multicolumn{14}{|c|}{\textbf{Mixed-type covariates}} \\
\midrule
\multirow{2}{*}{\shortstack{$375$ \\ $(125)$}} 
& UR & 0.158 & 0.146 & -0.112 & 0.353 & 0.895 
      & 0.112 & 0.101 & -0.106 & 0.326 & 0.945 
      & 0.254 & 0.960 \\
& LR & 0.136 & 0.126 &  0.100 & 0.351 & 0.945 
      & 0.136 & 0.126 &  0.101 & 0.320 & 0.910 
      & 0.252 & 0.955 \\
\cmidrule{1-14}
\multirow{2}{*}{\shortstack{$750$ \\ $(250)$}} 
& UR & 0.062 & 0.057 & -0.074 & 0.237 & 0.940 
      & 0.053 & 0.048 & -0.071 & 0.226 & 0.935 
      & 0.303 & 0.975 \\
& LR & 0.065 & 0.055 &  0.100 & 0.233 & 0.900 
      & 0.060 & 0.055 &  0.070 & 0.224 & 0.920 
      & 0.303 & 0.980 \\
\cmidrule{1-14}
\multirow{2}{*}{\shortstack{$1500$ \\ $(500)$}} 
& UR & 0.041 & 0.036 & -0.070 & 0.163 & 0.840 
      & 0.035 & 0.028 & -0.082 & 0.159 & 0.905 
      & 0.173 & 0.940 \\
& LR & 0.030 & 0.024 &  0.072 & 0.158 & 0.925 
      & 0.029 & 0.024 &  0.070 & 0.155 & 0.920 
      & 0.168 & 0.955 \\
\bottomrule
\end{tabular}
\end{adjustbox}
\begin{flushleft}
\textit{Note.} UR = unlikely responders; LR = likely responders; 
$n$ = total sample size per simulation run; 
$n_{\text{design}}$ = sample size used to train the model in the design stage.  
“XGBoost PBS” and “BART PBS”: XGBoost or BART was used to estimate PBS for subgrouping.
True treatment effects: Gaussian, $\Delta(\mathrm{UR}) = -0.704$, $\Delta(\mathrm{LR}) = 0.707$;  
Mixed-type, $\Delta(\mathrm{UR}) = -0.264$, $\Delta(\mathrm{LR}) = 0.261$.  
\end{flushleft}
\end{table*}

We conducted sensitivity analyses to assess robustness of the proposed method under several departures from the main simulation setup (see Supplementary Section~B). Specifically, we examined (i) misspecification due to unmeasured predictors (Tables~S1), (ii) alternative modeling choices in the second stage (Table~S2), (iii) data-generating mechanisms with near-perfect correlation between prognosis and treatment effect (Table~S3), and (iv) the consistency of results across different numbers of simulation iterations (Table~S4). Across these scenarios, results were consistent with the primary findings, showing small bias and near-nominal coverage, confirming that our proposed estimator is generally robust.

\section{A COVID-19 randomized clinical trial Example}

\subsection{Data Description}\label{data}

We apply the proposed two-stage sampling-based approach to evaluating treatment effects to data from $n=2341$ patients in the COMPILE COVID-19 clinical trial. The study evaluates the effectiveness of the COVID-19 convalescent plasma (CCP) treatment for hospitalized COVID-19 patients not on mechanical ventilation at the time of randomization \cite{troxel2022association}.
The primary outcome is a binary outcome derived from the World Health Organization (WHO) 11-point clinical scale, which assesses COVID-19 severity with values ranging from 0 (no infection) to 10 (death) \cite{marshall2020minimal}. (see Section C in the Supplemental Material.) 
Specifically, we consider four binary outcomes: 
(1) ventilation or death, defined as an indicator for a WHO score $\geq 7$ measured at 14 ± 1 days post-randomization (the primary outcome); 
(2) hospitalization, ventilation, or death, defined as an indicator for a WHO score $\geq 4$ measured at 28 ± 2 days post-randomization (secondary outcome); 
(3) ventilation or death, defined as an indicator for a WHO score $\geq 7$ measured at 28 ± 2 days post-randomization (secondary outcome); 
and (4) death, defined as an indicator for a WHO score of 10 at 28 ± 2 days post-randomization (secondary outcome).  Park et al.\cite{park2022development} developed a Treatment Benefit Index to identify hospitalized COVID-19 patients who may benefit from CCP,  however, their method ignores uncertainty in estimating data-driven model-based benefit groups. The treatment variable is a binary indicator (1 for CCP treatment, 0 for control).
%\st{The current paper pre-selected features in the COMPILE COVID-19 clinical trial based on the work by Park et al.\cite{park2022development} and Wu et al.\cite{wu2024bayesian}.}
Baseline covariates used in our study are as follows:

\begin{itemize}

%\item Treatment indicator: A binary variable (1 for CCP treatment, 0 for control).

\item Pre-treatment patient characteristics: 
Age (mean (SD) of 60.3 (15.2) years); %complete cases
Sex (35.6\% female); %complete cases
WHO score at baseline (19.5\% score 4, 62.7\% score 5, 17.8\% score 6);  
Blood type (categorical: A, B, AB, or O);
Indicator for diabetes (33.6\% yes);
Indicater for cardiovascular disease (42.2\% yes); 
Indicator for pulmonary disease (11.6\% yes); 
Duration of symptoms before randomization (integer range: 1–5).

\item Confounders: Quarter during which the patient was enrolled (a categorical variable: Jan–Mar 2020, Apr–Jun 2020, Jul–Sep 2020, Oct–Dec 2020, Jan–Mar 2021); RCTs (sites) participating in COMPILE (an 8-level categorical variable representing RCT ID).

\end{itemize}

\subsection{Application Results}

We evaluated the performance of three methods: the ``naïve’’ method, the ``naïve’’ method with BART, and the Bayesian-corrected two-stage method. Complete cases yield a final sample of 2,287 patients. We used the primary outcome (the binary indicator for ventilation or worse on day 14) as the outcome in the PBS model in the ``design’’ stage, and will then evaluate the data-driven subgroup-specific treatement effects with respect to the four binary outcomes in the ``evaluation’’ stage.

To estimate the estimand in each subgroup, we implemented a Monte Carlo (MC) cross-validation (CV) procedure with 100 repetitions, as in Goligher et al. \cite{goligher2023heterogeneous}. 
MC–CV is employed in the application to mitigate the overfitting risk inherent to single-sample analyses, whereas the simulation study uses genuinely independent design and evaluation samples and therefore does not require cross-validation.
In each MC iteration, we followed the following steps:

\begin{itemize} 
\item Individuals with $T=1$ were randomly divided into two equal-sized subsets: one assigned to a CV-``design’’ set, while the other was combined with all untreated individuals to form a CV-``evaluation’’ set. 
\item The PBS model was trained on the CV-``design’’ set using pre-treatment characteristics and confounders as predictors, and using the primary outcome. The trained model was then applied to individuals in the CV-``evaluation’’ set to calculate the PBS, where we set the confounder variables at RCT ID = AA and enrollment quarter = Apr–Jun 2020 to standardize confounding effects in the PBS calculation. 

\item After categorizing each individual in the CV-``evaluation’’ set into subgroups based on the computed PBS, the OR for subgroup-specific treatment effects was estimated for each subgroup across four binary outcomes adjusted for RCT ID and enrollment quarter.
\end{itemize}

A final predicted estimand was obtained by taking the median point estimate and confidence interval across 100 MC replications \cite{goligher2023heterogeneous}. The reported metrics in Table \ref{tab3} are as follow: 
1) $\mbox{OR}$: exponentiated median of the log ORs; 
2) $\mbox{SE}_{logOR}$: median of the standard errors of the log ORs; and 
3)  $\mbox{95\% CI}$; exponentiated median of the lower and upper bounds of the 95\% confidence intervals of the log ORs.

\begin{figure}[h] 
\centering
\includegraphics[width=17cm]{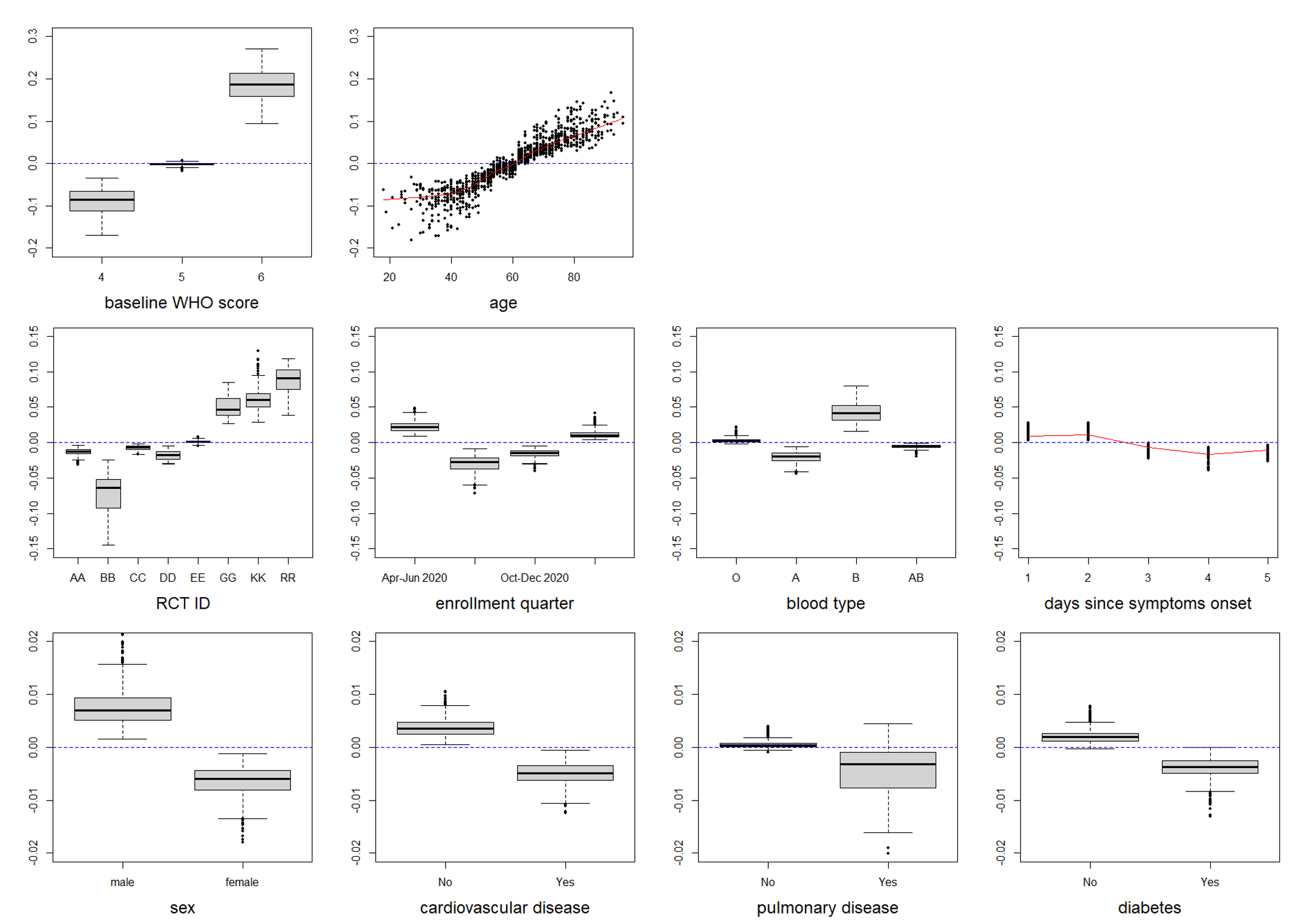}
\caption{Scatterplots of the feature values on the x-axis versus the corresponding SHAP values on the y-axis. 
The dashed horizontal line represents a SHAP value of zero, where points above the line indicate a positive contribution to the binary outcome of ventilation or death at day 14, while points below indicate a negative contribution. 
The range of SHAP values varies across the plots. 
Features in the first row, including baseline WHO score and age, exhibit the largest range (-0.2, 0.3). 
Features in the second row, including two clinical features (blood type, days since symptom onset) and the two confounders (RCT ID and enrollment quarter), have a smaller range (-0.15, 0.15). 
Features in the third row, including sex and preexisting conditions (cardiovascular disease, pulmonary disease, and diabetes), have the smallest range (-0.02, 0.02). 
More important predictors show greater variability along the y-axis.}
\label{shap}
\end{figure}

For the primary analysis, PBS corresponds to the predicted probability (risk) of ventilation or worse at day 14. 
In the analysis, we set thresholds at 0.1 and 0.2, categorizing individuals in the CV-``evaluation’’ set into three groups: 
if \( PBS \leq 0.1 \), then ``likely responder’’ (LR); 
if \( 0.1 < PBS \leq 0.2 \), then ``moderate responder’’ (MR); 
if \( PBS \geq 0.2 \), then ``unlikely responder’’ (UR).
Table \ref{tab3} presents the ORs and their corresponding 95\% confidence intervals for each subgroup, using various binary outcomes. 
For interpretation of PBS model, in Figure~\ref{shap}, we present predictor importance in the PBS model computed based on all individuals with $T=1$ using their SHapley Additive exPlanations (SHAP) values. 
Figure \ref{shap} indicates that the baseline WHO score was the most influential predictor. Among CCP-treated patients, a baseline WHO score of 6 (hospitalized; oxygen by noninvasive ventilation or high flow) increased the predicted probability of the outcome event by 0.185 on average, while a baseline WHO score of 5 (hospitalized; oxygen by mask or nasal prongs) showed no change. A baseline WHO score of 4 (hospitalized; no oxygen therapy) decreased the predicted probability by 0.09 on average.
Age was the second most important feature, exhibiting an almost linear relationship. Among CCP-treated patients, older age was associated with a higher predicted probability of the outcome event.
If a patient was hospitalized three or more days after symptom onset, they were predicted on average to have a lower probability of the outcome event. In contrast, patients hospitalized within two days of symptom onset,  with CCP treatment, were predicted to have a higher probability of the outcome event. 
Individuals with blood type A or AB and female individuals were, on average, more likely to be predicted as responders to CCP, though the contribution of sex and blood type was small.
More detailed discussion on variable importance is provided in Section D of the Supplemental Material. 

As shown in Table \ref{tab3}, the subgroup-specific treatment effect point estimates (the ``OR’’ columns) for the LR, MR, and UR subgroups across all four outcomes are similar when using the BART method and the XGBoost method in the ``design’’ stage. 
Regardless of the outcome evaluated, we observe a monotonic trend in OR estimates across the LR, IR, and UR subgroups. In the LR group, the odds of an unfavorable outcome comparing CCP treatment vs. control are the lowest, around 0.7. In the MR group, the odds remain somewhat lower, indicating a moderate treatment effect. However, in the UR group, the OR approaches 1, suggesting little to no treatment benefit in this subgroup. 
Among the outcomes evaluated, the CCP treatment provides the greatest benefit for reducing mortality at day 28. The estimated OR for mortality at day 28 is as low as 0.608, though the confidence intervals indicate that the results are not statistically significant. 
In terms of uncertainty quantification, the median standard error of the Bayesian-corrected two-stage method is consistently larger than that of the ``naïve’’ method across all subgroups and outcomes. This highlights that the naïve method underestimates the uncertainty from data-driven subgrouping. 
Note that the (log-scale) SE of the subgroup-specific treatment effect is not directly comparable between the BART and XGBoost methods, as these methods result in different subgroup sample sizes for the ``evaluation’’ set. When using XGBoost, the median (SD) sample sizes in the CV-``evaluation’’ set are 643 (187) for LR, 667.5 (167) for MR, and 326 (94) for UR. In contrast, BART trained on the CV-``design’’ set yields subgroup sizes of 778 (164) for LR, 490 (88) for MR, and 403.5 (108) for UR in the CV-``evaluation’’ set. 
The results reported on the log scale (i.e., log ORs) are provided in Section E of the Supplemental Material.

\begin{table*}[h]

\caption{Data Analysis Results for COMPILE COVID-19 clinical trial. Each CV-``evaluation’’ set consists of 1,692 samples. The ``naïve’’ method, using XGBoost trained on the CV-``design’’ set, yields median (SD) sample sizes of 643 (187) for likely-responders (LR), 667.5 (167) for moderate-responders (MR), and 326 (94) for unlikely-responders (UR). In the two-stage model, BART trained on the CV-``design’’ set produces sizes of 778 (164), 490 (88), and 403.5 (108) for LR, MR and UR, respectively.}\label{tab3}

\begin{adjustbox}{max width=\textwidth}
\begin{tabular}{|l| ccc | ccc cc|}
\toprule 
& \multicolumn{3}{c|}{\textbf{XGBoost} \ PBS}    & \multicolumn{5}{c|}{\textbf{BART} \ PBS }  \\
\cmidrule(lr){2-4} \cmidrule(lr){5-9}
& & \multicolumn{2}{c|}{\textbf{Naïve}} & & \multicolumn{2}{c}{\textbf{Naïve}} & \multicolumn{2}{c|}{\textbf{Corrected}} \\
\cmidrule(lr){3-4}
\cmidrule(lr){6-7}
\cmidrule(lr){8-9}
 & OR & $SE_{logOR}$ & 95\% CI 
& OR & $SE_{logOR}$ & 95\% CI  &  $ SE_{logOR}$ & 95\% CI \\
\midrule 
\multicolumn{9}{|c|}{Binary outcome: ventilation or worse at day 14} \\
\cmidrule{1-9}
\begin{tabular}{l}
LR 
\end{tabular} & 0.649 & 0.362  & (0.322, 1.254)
& 0.715 & 0.319 & (0.376, 1.301)  
& 0.368 & (0.348, 1.497) \\
\begin{tabular}{l} 
MR 
\end{tabular} & 0.955 & 0.245  & (0.595, 1.539) 
& 0.892 & 0.282 & (0.513, 1.570)  
& 0.426 & (0.386, 2.092) \\
\begin{tabular}{l} 
UR 
\end{tabular} & 0.918 & 0.259  & (0.566, 1.517) 
& 1.004 & 0.241 & (0.637, 1.593)  
& 0.290 & (0.566, 1.755) \\
\begin{tabular}{l} 
All 
\end{tabular} & 0.850 & 0.149  & (0.636, 1.138) 
& 0.861 & 0.148 & (0.644, 1.150)  
& 0.148 & (0.644, 1.150) \\
\midrule
\multicolumn{9}{|c|}{Binary outcome: hospitalization at day 28} \\
\cmidrule{1-9}
\begin{tabular}{l}
LR 
\end{tabular} & 0.695 & 0.316  & (0.382, 1.258)
& 0.736 & 0.269 & (0.434, 1.203)  
& 0.316 & (0.394, 1.341) \\
\begin{tabular}{l} 
MR 
\end{tabular} & 0.907 & 0.205  & (0.616, 1.335)
& 0.904 & 0.235 & (0.568, 1.443)  
& 0.356 & (0.446, 1.861) \\
\begin{tabular}{l} 
UR 
\end{tabular} & 0.919 & 0.248  & (0.569, 1.491)
& 0.997 & 0.224 & (0.646, 1.516)  
& 0.274 & (0.568, 1.657) \\
\begin{tabular}{l} 
All 
\end{tabular} & 0.842 & 0.129  & (0.654, 1.085)
& 0.854 & 0.128 & (0.664, 1.099)  
& 0.128 & (0.664, 1.099) \\
\midrule
\multicolumn{9}{|c|}{Binary outcome: ventilation or worse at day 28} \\
\cmidrule{1-9}
\begin{tabular}{l}
LR 
\end{tabular} & 0.700 & 0.356  & (0.366, 1.340)  
& 0.731 & 0.311 & (0.404, 1.362)  
& 0.364 & (0.364, 1.550) \\
\begin{tabular}{l} 
MR 
\end{tabular} & 0.867 & 0.245  & (0.537, 1.371)  
& 0.814 & 0.278 & (0.471, 1.405)  
& 0.420 & (0.352, 1.843) \\
\begin{tabular}{l} 
UR 
\end{tabular} & 0.992 & 0.255  & (0.607, 1.593)  
& 1.019 & 0.237 & (0.641, 1.615)  
& 0.284 & (0.585, 1.733) \\
\begin{tabular}{l} 
All 
\end{tabular} & 0.840 & 0.146  & (0.631, 1.119)  
& 0.852 & 0.145 & (0.641, 1.132)  
& 0.145 & (0.641, 1.132) \\
\midrule
\multicolumn{9}{|c|}{Binary outcome: mortality at day 28} \\
\cmidrule{1-9}
\begin{tabular}{l}
LR 
\end{tabular} & 0.608 & 0.400  & (0.289, 1.290)  
& 0.658 & 0.365 & (0.328, 1.288)  
& 0.409 & (0.302, 1.446) \\
\begin{tabular}{l} 
MR 
\end{tabular} & 0.759 & 0.291  & (0.439, 1.320)  
& 0.751 & 0.330 & (0.396, 1.436)  
& 0.495 & (0.279, 1.992) \\
\begin{tabular}{l} 
UR 
\end{tabular} & 0.953 & 0.272  & (0.561, 1.588)  
& 0.925 & 0.254 & (0.566, 1.536)  
& 0.296 & (0.518, 1.670) \\
\begin{tabular}{l} 
All 
\end{tabular} & 0.771 & 0.164  & (0.560, 1.063)  
& 0.778 & 0.164 & (0.565, 1.073)  
& 0.164 & (0.565, 1.073) \\
\bottomrule
\end{tabular}
\end{adjustbox}
\begin{flushleft}
\textit{Note.} LR: likely-responders; MR: moderate-responders; UR: unlikely-responders; All: all individuals (unstratified);   
“XGBoost PBS” and “BART PBS”: XGBoost or BART was used to estimate PBS for subgrouping.
OR = odds ratio; SE$_{\log OR}$ = standard error of the log odds ratio; CI = confidence interval.
\end{flushleft}
\end{table*}

We conducted additional analyses to evaluate the BART-based two-stage Bayesian model in the COMPILE COVID-19 trial (see Supplementary Section~F). First, diagnostic density and empirical cumulative distribution function (ECDF) plots of the estimated prognostic score showed good overlap across treatment arms, supporting the covariate-overlap assumption (Figure~S2). Second, sensitivity analyses varying the allocation of sample sizes between CV-``design'' and CV-``evaluation'' subsets yielded stable subgroup proportions and consistent treatment effect estimates (Tables~S6–S7). Third, posterior predictive checks confirmed adequate model fit, with observed event rates lying well within the posterior predictive distribution (Figure~S3). These results reinforce the robustness of our findings in the primary analysis. Finally, Supplementary Section~F4 provides a single, explicit rule for subgroup assignment, together with a measure of classification confidence, to aid clinical interpretability.

In addition to applying to ``likely responders’’  (absolute benefit exceeding a threshold) subgroups, we also apply our Bayesian two-stage framework to subgroups stratified by predicted heterogeneous treatment effects (relative treatment effects compared to control). The results are provided in Supplemental Section G.

\section{Discussion}

This article formalizes a two-stage Bayesian sampling-based procedure to evaluate data-driven subgroup-specific treatment effects estimates while accounting for uncertainty in the design stage. To prevent double-dipping, where the same data is used both to develop the model and to evaluate its performance, potentially leading to overly optimistic results, we partition the RCT data into separate ``design’’ and ``evaluation’’ sets. While this reduces statistical power, it is necessary given the constraint of having only a single dataset. 
An alternative approach in the design stage is to use bootstrapping for uncertainty quantification; however, bootstrapping is generally limited when dealing with complex hierarchical structures. In contrast, Bayesian posterior distributions provide a more natural framework to quantify uncertainty. 

Beyond statistical considerations, the LR framework can have clinical and regulatory value. In trials with null or inconclusive average effects, LR analyses may suggest that a treatment benefits a subset of patients; in trials with positive averages, they can help characterize where responses are stronger. Within this framework, our Bayesian approach explicitly propagates uncertainty from subgroup identification to subgroup-specific inference, reducing the risk of overconfident conclusions. LR analyses also connect to enrichment in trial design. U.S. FDA guidance describes enrichment as prospectively selecting patients more likely to respond than the broader target population, thereby improving power \cite{FDA_EnrichmentGuidance}. LR subgroups identified within an RCT offer a data-driven input to such enrichment, because potential responders are characterized using baseline features.
We note that LR subgrouping based on prognosis under treatment $s(x) = E[Y\mid T=1,x]$
does not, in general, guarantee enrichment for individuals with the largest incremental treatment advantage
$\Delta(x) = E[Y  \mid T=1,x] - E[Y \mid T=0,x]$; however, even when the alignment between $s(x)$ and $\Delta(x)$ is weak or negative,
our two-stage procedure continues to provide well-calibrated uncertainty quantification for treatment effects within the resulting clinically defined (rule-based) subgroup (see Supplementary Section B3).

For future applications of this method in RCTs, several practical recommendations may improve implementation. 
First, since the prognostic score is derived solely from baseline covariates, subgrouping is treatment-blind and thus comparable across randomized arms. When diagnostic plots indicate limited common support between treatment arms, prespecified remedies (e.g., trimming or restricting estimation to overlapping regions) should be applied to ensure that subgroup contrasts remain interpretable. 
Second, the clinical threshold defining likely responders is necessarily context-dependent; a simple ``threshold sweep'' (reporting LR prevalence and effects over a small grid of plausible cutoffs, as in our LR/MR/UR illustration) provides a robustness check and aids clinical elicitation. 
Finally, when a single rule for subgroup assignment is desired,  patients can be classified using the posterior mean prognostic scores, accompanied by the posterior probability of LR as a measure of classification confidence (see Supplement Section~F4 for the application).

Although the method is developed in the context of an RCT, it can be extended to more general settings. In early-phase or pilot studies where only single-arm data are available, the ``design’’ stage can rely on treated patients alone to develop a biomarker-based score, which can then be used to define subgroups in a subsequent, independent trial. The approach is also applicable to other model-based, data-driven subgroup analyses, such as those based on stratification by heterogeneous treatment effects \cite{thanassoulis2016individualized,lee2017coronary}.

The proposed two-stage framework provides a principled approach for estimating subgroup-specific treatment effects while accounting for uncertainty in the subgrouping process. By integrating flexible nonparametric Bayesian models with subsequent statistical inference, the method enhances the reliability of data-driven subgroup analyses to support more informed clinical decision-making.

\section*{Acknowledgments}
The research was supported by the National Institutes of 
Health under Awards  R01AA030045 and UL1TR001445.

\clearpage
\begin{center}
{\Large\bfseries Supplemental Material}
\end{center}
\vspace{1em}

\setcounter{table}{0}
\renewcommand{\thetable}{S\arabic{table}}
\setcounter{figure}{0}
\renewcommand{\thefigure}{S\arabic{figure}}
\setcounter{equation}{0}
\renewcommand{\theequation}{S\arabic{equation}}

\section*{A. Introduction of Tree-based Models}

\subsection*{A1.  Extreme gradient boosting}

The XGBoost method \cite{chen2016xgboost} is based on the concept of gradient boosting , a machine learning technique that enhances weak learners into strong ones. Friedman et al.\cite{friedman2000additive} later adapted boosting to a forward stagewise additive model that minimizes a loss function and brought forth a highly adaptable algorithm, gradient boosting machines (GBM). Specifically, a GBM uses $m$ additive functions to predict the outcome,

\begin{equation}
    \Pr(Y = 1 | X = x) = \phi(x) = \sum_{j=1}^{m} f_j(x),
\label{add}
\end{equation}

where $f = w_{q(x)}$ is a tree structure with decision rules $q$ that map an observation $\mathbf{x}$ to a terminal node, and $w$ represents the weights associated with terminal nodes. Each tree contains a continuous score on each of the terminal nodes. For a given $\mathbf{x}$, we will drop $\mathbf{x}$ down each of the $m$ boosted trees until it hits a terminal node of each tree, and calculate the final prediction for $\mathbf{x}$ by summing up the scores (given by $w$) in the corresponding terminal nodes across the $m$ trees, that is, $\sum_{j=1}^{m} f_j(\mathbf{x})$ in equation (\ref{add}). 

The idea of gradient boosting is to minimize the loss function $\mathcal{L}(\phi) = \sum_{i} l(y_i, \hat{y}_i)$, where $\hat{y}_i$ and $y_i$ are, respectively, the predicted and observed outcome for the $i$th individual in the data. 
In this study, we used the R package \texttt{xgboost} to fit XGBoost. For regression, loss function is root mean squared error (RMSE), and for classification we selected cross-entropy loss. 

The XGBoost algorithm adds a term $\Omega(f_j)$ to the loss function $\mathcal{L}(\phi)$ of the traditional gradient tree boosting to penalize model complexity. The revised loss function for XGBoost is

\begin{equation}
    \mathcal{L}(\phi) = \sum_{i=1}^{n} l(y_i, \hat{y}_i) + \sum_{j=1}^{m} \Omega(f_j),
\end{equation}

where $\Omega(f) = \gamma T + \frac{1}{2} \lambda ||w||^2$, $T$ is the number of terminal nodes in a tree, $||w||$ represents the $\ell_2$ norm of terminal node weights, $\gamma$ and $\lambda$ are tuning parameters governing further partitioning of the predictor space. The penalty term $\Omega(f)$ helps to smooth the learned weights $w$ to prevent overfitting. 

XGBoost further mitigates overfitting using shrinkage and column subsampling. Shrinkage scales newly added weights after each boosting step, similar to traditional gradient tree boosting. Column subsampling, inspired by the Random Forest algorithm \cite{breiman2001random}, selects a random subset of predictors for splitting at each step, enhancing model robustness.\cite{chen2016xgboost}

\subsection*{A2.  Bayesian additive regression trees}

The main idea of BART is to use ``weak learners" (i.e., regression trees) added together to provide a clearer picture of the whole and a more powerful predictive model. Tree-based regression models are attractive because they have the ability to flexibly model additive, interaction, and nonlinear effects of predictors on the response $Y$. Let $\mathbf{x} = (x_1, \dots, x_p)^\top$ denote the predictors, where $p$ is the number of predictors. Let $\{T_j, M_j\}_{j=1}^{m}$ be a set of tree models, where $T_j$ denotes the $j$th tree, and $M_j$ denotes the means from the terminal nodes of the $j$th tree.

For this description, we assume the response

\begin{equation}
Y = f(\mathbf{x}) + \epsilon, \quad \text{where } \epsilon \sim N(0, \sigma^2). 
\label{eq2}
\end{equation}

The BART model for the mean function $f(\mathbf{x})$ can be expressed as:

\begin{equation}
    f(\mathbf{x}) = g(\mathbf{x}, T_1, M_1) + g(\mathbf{x}, T_2, M_2) + \dots + g(\mathbf{x}, T_m, M_m),
\label{eq.sum}
\end{equation}

$g(\mathbf{x}, T_j, M_j) = \mu_{jl}$ if $\mathbf{x} \in A_{jl}$ (i.e., constant over elements of the partition $A_{jl}$), where $A_{jl}$ is the $l$th partition in the $j$th tree $T_j$, and $M_j$ is the collection of ``mean'' parameters $\{\mu_{jl}\}$ associated with the leaves of the $j$th tree $T_j$. Define there are $b_j$ partitions in the $j$th tree, then $l=1 \cdots b_j$.

To elaborate more on (\ref{eq.sum}), for each non-terminal node of the tree, there is a "splitting rule" taking the form $x_s < c$, which consists of the "splitting predictor" $x_s \in \mathbb{R}$ and the "splitting value" $c \in \mathbb{R}$. These will be determined from the data along with the tree structure $T_j$ and the associated parameters $M_j$. 
In this study, we used the R package \texttt{BART} to fit BART.
Next we illustrate on the prior specification and posterior sampling of function \texttt{wbart}.

\subsubsection*{Prior Distribution Specification for the BART Model}

The BART model is a Bayesian probability model, requiring prior distribution specification. Chipman et al. \cite{chipman2010bart} simplified this step using independence assumptions:

\begin{equation}
    P((T_1,M_1), \dots, (T_m,M_m), \sigma^2) = P(T_1, \dots, T_m) P(M_1, \dots, M_m | T_1, \dots, T_m) P(\sigma^2),
\label{eq.multi}
\end{equation}

where

$$
\begin{aligned}
    P(T_1, \dots, T_m) &= \prod_{j=1}^{m} P(T_j), \\
    P(M_1, \dots, M_m | T_1, \dots, T_m) &= \prod_{j=1}^{m} P(M_j | T_j), \\
    P(M_j | T_j) &= \prod_{l=1}^{b_j} P(\mu_{jl} | T_j) 
\end{aligned}
$$

For the number of trees in model (\ref{eq2}), we follow the default specification in \texttt{BART}, setting $m = 200$. 

For $T_j$, the prior probability that nodes at depth $d$ are non-terminal is set as $\alpha(1 + d)^{-\beta}$ with base $\alpha = 0.95$ and power $\beta = 2$ \cite{chipman2010bart}, encouraging smaller trees. 
Under this distribution, tree size of $\{1, 2, 3, 4, 5\}$ has a prior probability of $\{0.05, 0.55, 0.28, 0.09, 0.03\}$, respectively.

The prior on each tree is given as $\mu_{j} \sim N(\mu_\mu/ m, \sigma^2_\mu)$, where the expectation $\mu_\mu$ is picked to be the range center $(Y_{\min} + Y_{\max})/2$ in the training set. The variance hyperparameter $\sigma^2_\mu$ is empirically chosen such that $k \sqrt{m} \sigma_\mu = (Y_{\max} - Y_{\min})/2$, with $k=2$ by default.

The prior on error variance $\sigma^2$ follows an inverse gamma distribution, $\sigma^2 \sim \text{InvGamma}(\nu/2, \nu\lambda/2)$, where $\lambda$ is determined from the data ensuring a 90\% prior probability that the RMSE is below that of an OLS regression, and $\nu=3$ by default.

\subsubsection*{Posterior Sampling of the BART Model}

Given the prior and the observed data, posterior sampling in BART models is carried out via Markov chain Monte Carlo (MCMC) based on "Bayesian back-fitting" \cite{hastie2000bayesian}, which is an application  of the "simple" Gibbs sampler. Specifically, the $j$th tree $(T_j, M_j)$ is fit iteratively, holding all other $m-1$ trees constant. For each iteration, given $\{T_j\}$ 
and $\{M_j\}$ and based on the model residual $R_i = Y_i - \sum_{j=1}^{m} g(x_i, T_j, M_j)$,
we draw a sample from the posterior of $\sigma^2$ ("\textit{Step 1}"). Then, we define the $j$th partial residual response

\begin{equation}
    R_{ij} = Y_i - \sum_{j' \neq j} g(x_i, T_{j'}, M_{j'}), \quad i = 1, \dots, n.
\label{eq.res}
\end{equation}

and use single-tree MCMC updates, treating the partial residuals $R_{ij}$ as the response data to fit the 
$j$th tree ("\textit{Step 2}"). These \textit{Steps 1 and 2} constitute a Metropolis-within-Gibbs sampler for 
posterior sampling of BART \eqref{eq.sum}:

\begin{enumerate}
    \item $\sigma^2 | \{T_j\}, \{M_j\}$;
    \item $T_j, M_j | \{T_i\}_{i \neq j}, \{M_i\}_{i \neq j}, \sigma^2$;
\end{enumerate}

where, within each $j$th block of \textit{Step 2}, we compute the Metropolis ratio:

$$
    r = \frac{P(T^*_j \to T_j)}{P(T_j \to T^*_j)} 
    \times \frac{P(R_j | T^*_j, \sigma^2)}{P(R_j | T_j, \sigma^2)} 
    \times \frac{P(T^*_j)}{P(T_j)}
$$

where $T^*_j$ is the proposal tree, and 
$R_j \in \mathbb{R}^n$ is the vector of partial residuals defined based on \eqref{eq.res}. 
We accept the proposal tree if a draw from a standard uniform distribution is less than the value of $r$. Then, given $T_j$, 
we sample the leaf parameters in $M_j$ from its posterior distribution which is conjugate normal with the mean 
being a weighted combination of the likelihood and prior parameters.

For example, the \texttt{wbart}'s default number of burn-in Gibbs samples $(= 100)$ can be used. 
However, to approximate the posterior fit of model \eqref{eq.sum}, we used 500 burn-in samples and kept 100 samples post-burn-in in both simulation and real data analysis.
Let $f^r$ denote the 
$r$th draw of $f$. Averaging the $f^r(x_i)$ values over $i$ with $r$ fixed, the resulting values are a Monte Carlo approximation to the posterior distribution of the expected response for the associated population. 
This creates posterior distributions for individual level means.

%\pagebreak
\clearpage

\section*{B. Simulation-Based Sensitivity Analysis}

\subsection*{B1. Sensitivity to Unmeasured Predictors}

To assess sensitivity to model misspecification, we conducted additional simulations in which the true outcome model depended on more covariates than were available for estimation, with the omitted predictors having modest but nontrivial effects. Specifically, we considered two settings: one where outcomes depended on 12 covariates but estimation used only the first 10, and another where outcomes depended on 15 covariates but estimation again used only the first 10.  
\bigbreak
\textbf{12-dimensional setting.}  
$X_i \sim N(0,I_{12})$, $f(\boldsymbol{x}_i, T_i) = \boldsymbol{\mu}^\top \boldsymbol{x}_i + g(\boldsymbol{x}_i) \cdot T_i,$ with  
\[
\boldsymbol{\mu} = [0.5,\,1,\,0.75,\,1,\,0.5,\,-0.5,\,-1,\,-0.75,\,-1,\,-0.5,\,0.1,\,-0.1],
\]    
\[
g(\boldsymbol{x}_i) = \tfrac{\sin(\pi x_{i1})}{5} - \tfrac{\sin(\pi x_{i2})}{5} + \tfrac{x_{i3}^2}{5} - \tfrac{x_{i4}^2}{5} - \tfrac{x_{i5}x_{i6}}{10} + x_{i7} - x_{i8} + \tfrac{\exp(x_{i9})}{5} - \tfrac{\exp(x_{i10})}{5} - \tfrac{|x_{i11}|}{10} - \tfrac{|x_{i12}|}{20}.
\]  
Estimation used only the first 10 covariates.  
\bigbreak
\textbf{15-dimensional setting.}  
$X_i \sim N(0,I_{15})$, $f(\boldsymbol{x}_i, T_i) = \boldsymbol{\mu}^\top \boldsymbol{x}_i + g(\boldsymbol{x}_i) \cdot T_i,$ with  
\[
\boldsymbol{\mu} = [0.5,\,1,\,0.75,\,1,\,0.5,\,-0.5,\,-1,\,-0.75,\,-1,\,-0.5,\,0.2,\,-0.2,\,0.1,\,-0.1,\,-0.05],
\]    
\[
g(\boldsymbol{x}_i) = \tfrac{\sin(\pi x_{i1})}{5} - \tfrac{\sin(\pi x_{i2})}{5} + \tfrac{x_{i3}^2}{5} - \tfrac{x_{i4}^2}{5} - \tfrac{x_{i5}x_{i6}}{10} + x_{i7} - x_{i8} + \tfrac{\exp(x_{i9})}{5} - \tfrac{\exp(x_{i10})}{5} - \tfrac{x_{i11}+x_{i12}+x_{i13}+x_{i14}+x_{i15}}{20}.
\]  
Again, estimation used only the first 10 covariates. 
\bigbreak
Results are summarized in Table~\ref{tab:misspec}. Across a range of sample sizes, the subgroup-specific treatment effect estimates remained decent, with bias close to zero and coverage near nominal levels. This indicates that the corrected estimator is robust to modest model misspecification, as the additional uncertainty is absorbed by between-design variability. In other words, while subgroup classification may be imperfect when predictors are unmeasured, the downstream causal effect estimates within identified subgroups remain valid.

\begin{table}[htbp]
\centering
\begin{threeparttable}
\caption{Simulation results for binary outcomes under covariate misspecification. 
Outcomes were generated with 12 or 15 covariates, but estimation used only the first 10.}
\label{tab:misspec}

\begin{tabular}{l l ccccc ccccc}
\toprule 
\multirow{3}{*}[\dimexpr 2ex]{\shortstack{\textbf{\(\text{n}_{\text{eval}}\)} \\ (\(\text{n}_{\text{design}}\))}} 
& & \multicolumn{5}{c}{\shortstack{\textbf{Generated with 12 covariates} \\ \textbf{Estimated with 10}}} 
& \multicolumn{5}{c}{\shortstack{\textbf{Generated with 15 covariates} \\ \textbf{Estimated with 10}}} \\
\cmidrule(lr){3-7}\cmidrule(lr){8-12}
& & MSE & Var & Bias & $\overline{SE}$ & Covg & MSE & Var & Bias & $\overline{SE}$ & Covg \\
\midrule
\multirow{2}{*}{375 (125)} 
& UR & 0.183 & 0.145 & -0.195 & 0.434 & 0.940 & 0.181 & 0.154 & -0.164 & 0.439 & 0.965 \\
& LR & 0.192 & 0.175 &  0.130 & 0.438 & 0.940 & 0.202 & 0.192 &  0.099 & 0.436 & 0.925 \\
\midrule
\multirow{2}{*}{750 (250)} 
& UR & 0.082 & 0.078 & -0.058 & 0.306 & 0.975 & 0.104 & 0.091 & -0.114 & 0.306 & 0.960 \\
& LR & 0.087 & 0.083 &  0.063 & 0.309 & 0.945 & 0.096 & 0.078 &  0.133 & 0.302 & 0.940 \\
\midrule
\multirow{2}{*}{1500 (500)} 
& UR & 0.048 & 0.043 & -0.077 & 0.215 & 0.945 & 0.051 & 0.037 & -0.118 & 0.215 & 0.930 \\
& LR & 0.047 & 0.041 &  0.078 & 0.214 & 0.935 & 0.048 & 0.043 &  0.071 & 0.214 & 0.960 \\
\bottomrule
\end{tabular}

\begin{tablenotes}[flushleft]
\item Abbreviations: UR, unlikely-responders; LR, likely-responders. $\text{n}_{\text{eval}}$: evaluation sample size; $\text{n}_{\text{design}}$: design sample size used to train the model.
\end{tablenotes}
\end{threeparttable}
\end{table}

\pagebreak

\subsection*{B2. Sensitivity to Second-Stage Modeling Choice}

Besides standard GLM, we explored Bayesian GLM in the second stage. 
The Bayesian and standard GLM approaches produced similar estimates (Table~\ref{bayesGLM}). 
Both approaches begin with BART in the design stage to obtain $K$ posterior draws of the PBS. As described in the main text (Section \textit{Computational procedure for Two-Stage Sampling}), the standard GLM then uses Rubin's rule to combine uncertainty across these $K$ draws. 
In the Bayesian GLM specification, for each draw $\boldsymbol{\nu}_k$, a Bayesian GLM is fit in the second stage to generate $M$ posterior draws of $\Delta_k \mid \boldsymbol{\nu}_k$. The resulting $K \times M$ draws are then combined to compute posterior means and variances of the subgroup-specific treatment effect estimates, with credible intervals obtained from posterior quantiles.

\begin{table*}[h]
\centering
\begin{threeparttable}
\caption{Comparison of Bayesian GLM and standard GLM (posterior-design-averaged) for binary outcomes. Results show show small differences in accuracy and coverage across scenarios.}
\label{bayesGLM}

\begin{tabular}{|c c | ccccc | ccccc |}
\toprule 
\multirow{3}{*}{\shortstack{\textbf{\(\text{n}_{\text{eval}}\)} \\ \((\text{n}_{\text{design}})\)}} & & \multicolumn{5}{c|}{\textbf{Bayesian GLM}} & \multicolumn{5}{c|}{\textbf{Posterior-design-averaged GLM}} \\
\cmidrule[1pt](r){3-7}
\cmidrule[1pt](r){8-12}
& & MSE & Var & Bias & $\overline{SE}$ & Covrg & MSE & Var & Bias & $\overline{SE}$ & Covrg \\
\midrule
\multirow{2}{*}{\shortstack{$375$ \\ $(125)$}} & UR
& 0.219 & 0.198 & -0.145 & 0.441 & 0.940 
& 0.195 & 0.155 & -0.199 & 0.439 & 0.940 \\
& LR 
& 0.222 & 0.180 &  0.203 & 0.434 & 0.935 
& 0.240 & 0.212 &  0.167 & 0.442 & 0.915 \\
\midrule
\multirow{2}{*}{\shortstack{$750$ \\ $(250)$}} & UR 
& 0.089 & 0.082 & -0.085 & 0.308 & 0.960 
& 0.100 & 0.092 & -0.089 & 0.311 & 0.965 \\
& LR
& 0.090 & 0.078 &  0.108 & 0.305 & 0.950 
& 0.085 & 0.074 &  0.106 & 0.303 & 0.950 \\
\midrule
\multirow{2}{*}{\shortstack{$1500$ \\ $(500)$}} & UR 
& 0.043 & 0.035 & -0.085 & 0.216 & 0.970 
& 0.042 & 0.036 & -0.080 & 0.215 & 0.965 \\
& LR
& 0.047 & 0.043 &  0.069 & 0.215 & 0.945 
& 0.044 & 0.040 &  0.068 & 0.215 & 0.950 \\
\bottomrule
\end{tabular}

\begin{tablenotes}[flushleft]
\item Abbreviations: UR, unlikely-responders; LR, likely-responders. $\text{n}_{\text{eval}}$: evaluation sample size; $\text{n}_{\text{design}}$: design sample size used to train the model. 
\end{tablenotes}
\end{threeparttable}
\end{table*}

\subsection*{B3. Sensitivity to Alignment between $E[Y\mid T,x]$ and $E[Y\mid T,x]-E[Y\mid C,x]$}

The likely responder (LR) subgroup is defined by a clinically prespecified threshold on
$s(x)=E[Y\mid T=1,x]$ and, in general, does not coincide with selection based on the individual treatment effect
$\Delta(x)=E[Y\mid T=1,x]-E[Y\mid T=0,x]$.
To examine the behavior of the proposed two-stage Bayesian procedure under different relationships between prognosis and treatment effect, we conducted sensitivity analyses under data-generating mechanisms with varying degrees of alignment between $s(x)$ and $\Delta(x)$.

Across all scenarios, covariates were generated as $X_i\sim N(0,I_{10})$, and the LR subgroup was defined using a common clinical criterion with $\texttt{minCond}=0$, corresponding to $s(x)\ge 0$.

For the binary outcome, responses were generated from 
$\text{logit}\{E(Y\mid T,X)\} =
\boldsymbol{\mu}_0^\top X + \boldsymbol{\mu}_1^\top X \cdot T$.

For the continuous outcome, responses were generated from
$E(Y\mid T,X) =
\boldsymbol{\mu}_0^\top X + \boldsymbol{\mu}_1^\top X \cdot T$.

Different choices of $(\boldsymbol{\mu}_0,\boldsymbol{\mu}_1)$ were used to induce varying degrees of alignment between $s(x)$ and $\Delta(x)$, as described below.

\paragraph{Strong positive alignment.}
To induce strong positive alignment between $s(x)$ and $\Delta(x)$ (Table~\ref{tab:correlation}), we set 
$\boldsymbol{\mu}_0 =
\begin{bmatrix}
1.2,\,1.0,\,0.8,\,0.6,\,0.4,\,0.2,\,-0.2,\,-0.4,\,-0.6,\,-0.8
\end{bmatrix}^\top$,\\
with
$\boldsymbol{\mu}_1 =
\begin{bmatrix}
1.08,\,0.90,\,0.72,\,0.54,\,0.36,\,0.18,\,-0.18,\,-0.36,\,-0.54,\,-0.72
\end{bmatrix}^\top
\quad \text{(binary outcome)}$,\\
and
$\boldsymbol{\mu}_1 =
\begin{bmatrix}
0.36,\,0.30,\,0.24,\,0.18,\,0.12,\,0.06,\,-0.06,\,-0.12,\,-0.18,\,-0.24
\end{bmatrix}^\top
\quad \text{(continuous outcome)}$.

\paragraph{Weak or moderately negative alignment.}
To induce weak or moderately negative alignment between $s(x)$ and $\Delta(x)$ (Table~\ref{tab:negative_cor}), we retained the same $\boldsymbol{\mu}_0$ and set\\
$\boldsymbol{\mu}_1 =
\begin{bmatrix}
0.020,\,-0.255,\,-0.095,\,-0.015,\,0.081,\,-0.002,\,0.115,\,-0.109,\,0.000,\,0.203
\end{bmatrix}^\top
\quad \text{(binary outcome)}$,
and\\
$\boldsymbol{\mu}_1 =
\begin{bmatrix}
0.040,\,-0.509,\,-0.190,\,-0.030,\,0.162,\,-0.003,\,0.230,\,-0.218,\,-0.001,\,0.406
\end{bmatrix}^\top
\quad \text{(continuous outcome)}$.

Across both alignment regimes, the corrected two-stage procedure yields well-calibrated uncertainty for subgroup-specific treatment effects under the LR definition, with near-nominal empirical coverage across sample sizes and outcome types (Tables~\ref{tab:correlation}--\ref{tab:negative_cor}).

\begin{table*}[h]
\centering
\begin{threeparttable}
\caption{Simulation results for the corrected estimator under a data generative mechanism where prognosis and treatment effect are nearly perfectly correlated.}
\label{tab:correlation}

\begin{tabular}{|c c | ccccc | ccccc |}
\toprule 
\multirow{3}{*}{\shortstack{\textbf{\(\text{n}_{\text{eval}}\)} \\ \((\text{n}_{\text{design}})\)}} & & \multicolumn{5}{c|}{\textbf{Binary Outcome}} & \multicolumn{5}{c|}{\textbf{Continuous Outcome}} \\
\cmidrule[1pt](r){3-7}
\cmidrule[1pt](r){8-12}
& & MSE & Var & Bias & $\overline{SE}$ & Covrg & MSE & Var & Bias & $\overline{SE}$ & Covrg \\
\midrule
\multirow{2}{*}{375 (125)} 
& UR & 0.215 & 0.201 & -0.120 & 0.499 & 0.955 & 0.082 & 0.081 & -0.031 & 0.335 & 0.980 \\
& LR & 0.213 & 0.210 & \phantom{-}0.060 & 0.493 & 0.955 & 0.094 & 0.094 & \phantom{-}0.017 & 0.335 & 0.970 \\
\midrule
\multirow{2}{*}{750 (250)} 
& UR & 0.110 & 0.107 & -0.052 & 0.334 & 0.930 & 0.050 & 0.050 & -0.005 & 0.229 & 0.950 \\
& LR & 0.094 & 0.086 & \phantom{-}0.088 & 0.334 & 0.950 & 0.042 & 0.042 & \phantom{-}0.008 & 0.226 & 0.975 \\
\midrule
\multirow{2}{*}{1500 (500)} 
& UR & 0.057 & 0.051 & -0.072 & 0.234 & 0.945 & 0.027 & 0.027 & -0.019 & 0.155 & 0.940 \\
& LR & 0.056 & 0.048 & \phantom{-}0.091 & 0.232 & 0.940 & 0.022 & 0.022 & \phantom{-}0.018 & 0.155 & 0.950 \\
\bottomrule
\end{tabular}

\begin{tablenotes}[flushleft]
\item Abbreviations: UR, unlikely-responders; LR, likely-responders. $\text{n}_{\text{eval}}$: evaluation sample size; $\text{n}_{\text{design}}$: design-stage sample size.
\item Empirical correlation between $s(x)$ and $\Delta(x)$ in the full simulated population:
binary outcome Spearman $\rho=0.66$, Pearson $r=0.84$; continuous outcome Spearman $\rho=1$, Pearson $r=1$.
\end{tablenotes}
\end{threeparttable}
\end{table*}

\begin{table*}[h]
\centering
\begin{threeparttable}
\caption{Simulation results for the corrected estimator under a data generative mechanism where prognosis and treatment effect are moderately negatively correlated.}
\label{tab:negative_cor}

\begin{tabular}{|c c | ccccc | ccccc |}
\toprule 
\multirow{3}{*}{\shortstack{\textbf{\(\text{n}_{\text{eval}}\)} \\ \((\text{n}_{\text{design}})\)}} & & \multicolumn{5}{c|}{\textbf{Binary Outcome}} & \multicolumn{5}{c|}{\textbf{Continuous Outcome}} \\
\cmidrule[1pt](r){3-7}
\cmidrule[1pt](r){8-12}
& & MSE & Var & Bias & $\overline{SE}$ & Covrg & MSE & Var & Bias & $\overline{SE}$ & Covrg \\
\midrule
\multirow{2}{*}{375 (125)} 
& UR & 0.137 & 0.137 &  0.014 & 0.415 & 0.975 & 0.070 & 0.070 &  0.014 & 0.319 & 0.980 \\
& LR & 0.170 & 0.169 & -0.029 & 0.416 & 0.960 & 0.067 & 0.067 & -0.012 & 0.321 & 0.990 \\
\midrule
\multirow{2}{*}{750 (250)} 
& UR & 0.068 & 0.068 &  0.018 & 0.288 & 0.985 & 0.035 & 0.035 &  0.005 & 0.217 & 0.970 \\
& LR & 0.070 & 0.069 & -0.028 & 0.287 & 0.985 & 0.030 & 0.030 & -0.008 & 0.218 & 0.975 \\
\midrule
\multirow{2}{*}{1500 (500)} 
& UR & 0.034 & 0.033 &  0.012 & 0.201 & 0.975 & 0.017 & 0.017 & -0.006 & 0.149 & 0.975 \\
& LR & 0.030 & 0.030 & -0.001 & 0.201 & 0.975 & 0.018 & 0.018 & -0.014 & 0.149 & 0.975 \\
\bottomrule
\end{tabular}

\begin{tablenotes}[flushleft]
\item Abbreviations: UR, unlikely-responders; LR, likely-responders. $\text{n}_{\text{eval}}$: evaluation sample size; $\text{n}_{\text{design}}$: design-stage sample size.
\item Empirical correlation between $s(x)$ and $\Delta(x)$ in the full simulated population:
binary outcome Spearman $\rho=-0.17$, Pearson $r=-0.10$; continuous outcome Spearman $\rho=-0.17$, Pearson $r=-0.18$.
\end{tablenotes}
\end{threeparttable}
\end{table*}

\pagebreak

\subsection*{B4. Monte Carlo Error and Simulation Precision}

We quantified Monte Carlo (MC) error in our simulation setting to guarantee that observed differences in performance patterns between methods are robust and not attributable to random simulation variation.

For coverage, the Monte Carlo standard error (MCSE) of a proportion $\hat{p}$ based on $R=200$ independent iterations is
\[
\mathrm{MCSE}(\hat{p}) = \sqrt{\hat{p}(1-\hat{p})/R}.
\]

At $\hat{p}\approx0.95$ and $R{=}200$, MCSE $\approx 0.015$, which is small relative to the reported differences.

For bias, the MCSE of the estimated mean bias $\overline{b}$ across $R=200$ simulation replicates is
\[
\mathrm{MCSE}(\overline{b}) = \frac{\mathrm{SD}(b)}{\sqrt{R}},
\]

where $\mathrm{SD}(b)$ denotes the empirical standard deviation of individual bias estimates $b_r = \hat{\Delta}_r - \Delta$.
Based on our simulation results, $\mathrm{SD}(b){<}0.45$ at $n_{\mathrm{design}}=125$ (MCSE ${<}0.03$), $\mathrm{SD}(b){<}0.3$ at $n_{\mathrm{design}}=250$ (MCSE ${<}0.02$), and $\mathrm{SD}(b){<}0.2$ at $n_{\mathrm{design}}=500$ (MCSE ${<}0.015$).
These values indicate that the principal performance patterns reported in the main text are unlikely to be driven by MC noise.

As a robustness check, we repeated the binary-outcome (Gaussian covariates) simulation using $R=500$ iterations (Table~\ref{iterations}). Across all sample sizes, differences in estimated bias between the $R=200$ and $R=500$ runs were approximately $0.01$, except for the unlikely-responder (UR) subgroup at $n_{\mathrm{design}}=125$, where the difference was $0.07$, consistent with greater variability at the smallest training size.

\begin{table*}[h]
\centering
\begin{threeparttable}
\caption{Comparison of simulation results with $R=200$ vs $R=500$ iterations (binary outcomes, Gaussian covariates).}
\label{iterations}

\begin{tabular}{|c c | ccccc | ccccc |}
\toprule 
\multirow{3}{*}{\shortstack{\textbf{\(\text{n}_{\text{eval}}\)} \\ \((\text{n}_{\text{design}})\)}} & & \multicolumn{5}{c|}{$R=500$} & \multicolumn{5}{c|}{$R=200$} \\
\cmidrule[1pt](r){3-7}
\cmidrule[1pt](r){8-12}
& & MSE & Var & Bias & $\overline{SE}$ & Covrg & MSE & Var & Bias & $\overline{SE}$ & Covrg \\
\midrule
\multirow{2}{*}{\shortstack{$375$ \\ $(125)$}} & UR
& 0.220 & 0.204 & -0.126 & 0.443 & 0.940 
& 0.195 & 0.155 & -0.199 & 0.439 & 0.940 \\
& LR 
& 0.207 & 0.183 &  0.157 & 0.434 & 0.926 
& 0.240 & 0.212 &  0.167 & 0.442 & 0.915 \\
\midrule
\multirow{2}{*}{\shortstack{$750$ \\ $(250)$}} & UR 
& 0.101 & 0.091 & -0.103 & 0.308 & 0.954 
& 0.100 & 0.092 & -0.089 & 0.311 & 0.965 \\
& LR
& 0.092 & 0.078 &  0.115 & 0.303 & 0.946 
& 0.085 & 0.074 &  0.106 & 0.303 & 0.950 \\
\midrule
\multirow{2}{*}{\shortstack{$1500$ \\ $(500)$}} & UR 
& 0.045 & 0.040 & -0.067 & 0.217 & 0.954 
& 0.042 & 0.036 & -0.080 & 0.215 & 0.965 \\
& LR
& 0.044 & 0.039 &  0.071 & 0.215 & 0.950 
& 0.044 & 0.040 &  0.068 & 0.215 & 0.950 \\
\bottomrule
\end{tabular}

\begin{tablenotes}[flushleft]
\item Abbreviations: UR, unlikely-responders; LR, likely-responders. $\text{n}_{\text{eval}}$: evaluation sample size; $\text{n}_{\text{design}}$: design sample size used to train the model. 
\end{tablenotes}
\end{threeparttable}
\end{table*}

%\pagebreak
\clearpage

\section*{C. The World Health Organization (WHO) 11-point COVID-19 clinical status scale.}

\begin{table}[h]
    \centering
    \caption{Full WHO (World Health Organization) 11-point COVID-19 patient status scale. This WHO 11-scale ordinal outcome was measured at day 14 and day 28 post-treatment as well as at baseline.}\label{who}
    \renewcommand{\arraystretch}{1.3}
    \begin{tabular}{cl}
        \toprule
        \textbf{Score} & \textbf{Descriptor} \\
        \midrule
        0  & Uninfected; no viral RNA detected \\
        1  & Asymptomatic; viral RNA detected \\
        2  & Symptomatic: Independent \\
        3  & Symptomatic: assistance needed \\
        4  & Hospitalized; no oxygen therapy \\
        5  & Hospitalized; oxygen by mask or nasal prongs \\
        6  & Hospitalized; oxygen by Noninvasive ventilation or High flow \\
        7  & Intubation \& Mechanical ventilation, pO2/FIO2$\geq$150 or SpO2/FIO2$\geq$200 \\
        8  & Mechanical ventilation pO2/FIO2$<$150 (SpO2/FIO2$<$200) or vasopressors \\
        9  & Mechanical ventilation pO2/FIO2$<$150 and vasopressors, dialysis, or ECMO \\
        10 & Death \\
        \bottomrule
    \end{tabular}
\end{table}

%\pagebreak
\clearpage

\section*{D. Variable Importance in the BART Model for COMPILE COVID-19 Clinical Trial}

\subsection*{D1. Variable Inclusion Proportion}
\begin{figure}[h] 
\centering
\includegraphics[width=14cm]{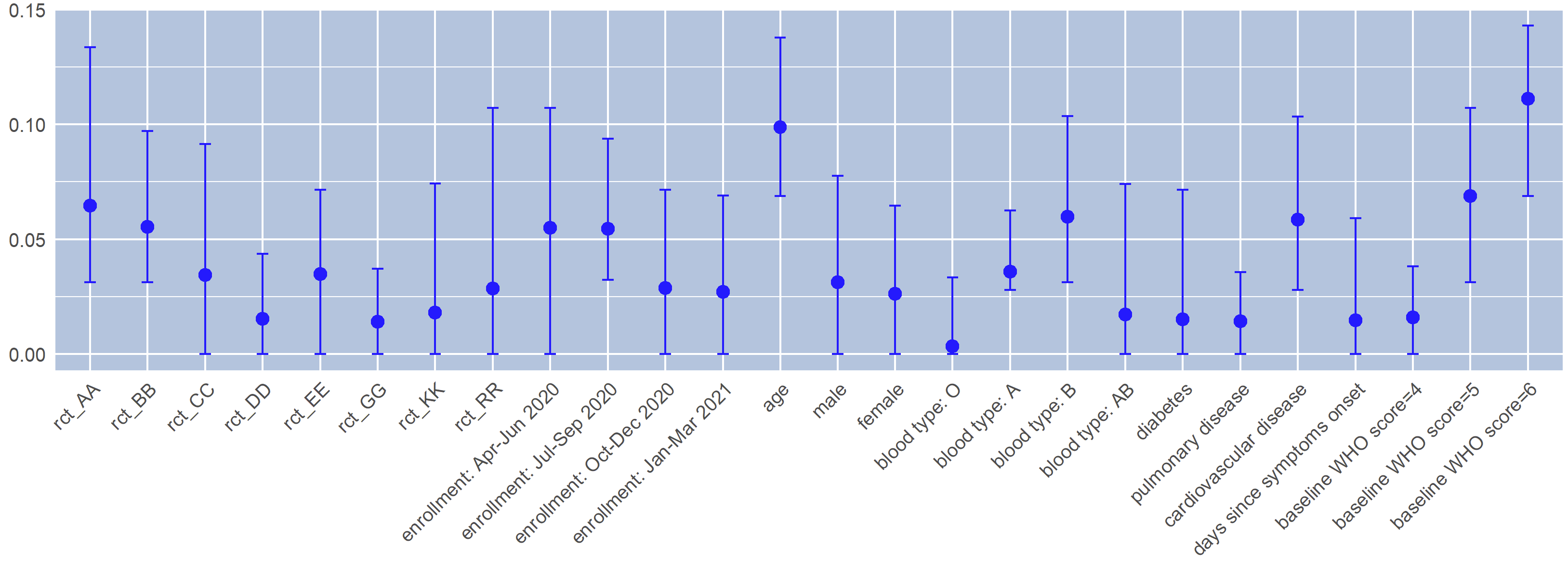}
\caption{Variable inclusion proportions. The dots mark the posterior means; the vertical error bars mark the 95\% credible intervals.}
\label{vip}
\end{figure}

In this section, we report results supplementary to Section 5.2 of the main manuscript, regarding the prognostic balancing score model developed in the design stage.
In the context of BART, variable importance is conveyed through the Variable Inclusion Proportion (VIP).\cite{bleich2014variable} The VIP represents the proportion of posterior MCMC samples in which a variable was included in a splitting rule in the sum-of-trees model, as described by Chipman et al.\cite{chipman2010bart}.
Figure \ref{vip} displays the VIPs from a BART model trained on all treated individuals. In order to sharpen our variable selection, we limited the model to 20 trees when computing the VIPs.

We observed that age and baseline WHO score exhibited a high proportion of inclusion in the model, with the lower bounds of their 95\% credible intervals exceeding 0.05. Furthermore, variables such as blood type, trial ID, enrollment quarter, and cardiovascular disease showed moderately high proportions of inclusion, with lower bounds of their 95\% credible intervals exceeding 0.025.

\subsection*{D2. SHapley Additive exPlanations values}

In addition to VIP, contributions of each candidate feature to outcome prediction were also quantified by SHapley Additive exPlanations (SHAP) values. 
% Specifically, a SHAP value of a feature captures the difference in the probability of an outcome event attributable to the difference in the value of a feature between patients.
SHAP values are derived from the principle that a feature's importance should be assessed by considering all possible combinations of features, where each feature may or may not be included in the model. This approach requires training separate models for every possible combination of features while keeping the hyperparameters and the training data constant.
Once these models are trained, SHAP values are computed by analyzing the differences in model outputs when including or excluding each specific feature. 
In our analysis, SHAP values were calculated using a BART model trained on all CCP treated individuals, with the binary outcome of ventilation or death at 14 ± 1 days post-randomization. The model is multivariate, adjusted for all covariates as well as confounders (RCT ID and enrollment quarter).
% however, since RCT ID and enrollment quarter are confounders, they are not included in the SHAP analysis.

For each feature, SHAP values were computed at the individual level to quantify its contribution to the outcome. 
The scatter plots in Figure 2 in the main text indicate how SHAP estimates for the LR group assignment relate to each pre-treatment characteristic. The plots are ordered by the variability of SHAP values (y-axis), arranged from highest to lowest across rows: top (row 1), middle (row 2), and bottom (row 3). 
To calculate SHAP values, we utilize function in the 'iml' R package.

As indicated in main text, we classified individuals with $PBS \leq 0.1$ as likely responders.
This implies that a lower probability of ventilation or death at day 14 under CCP treatment (i.e., a lower PBS) corresponds to a higher likelihood of being a responder.
The figure indicates that the baseline WHO score was the most influential predictor. Among CCP-treated patients, a baseline score of 6 (hospitalized; oxygen by NIV or high flow) increased the predicted probability of the outcome event by 0.185 on average, while a baseline score of 5 (hospitalized; oxygen by mask or nasal prongs) showed no change. A baseline score of 4 (hospitalized; no oxygen therapy) decreased the predicted probability by 0.09 on average.
Age was the second most important feature, exhibiting an almost linear relationship. Among CCP-treated patients, older age was associated with a higher predicted probability of the outcome event.
If a patient was hospitalized three or more days after symptom onset, they were predicted on average to have a lower probability of the outcome event. In contrast, patients hospitalized within two days of symptom onset, even with CCP treatment, were predicted to have a higher probability of the outcome event. Since all patients in SHAP analysis received CCP treatment, this difference may be attributed to baseline severity of illness, suggesting potential selection bias.
Individuals with blood type A or AB and female individuals were, on average, more likely to be predicted as responders to CCP, though the contribution of sex and blood type was small.
 
Patients with preexisting conditions—cardiovascular disease, pulmonary disease, and diabetes—were, on average, predicted to have a lower probability of the outcome event compared to those without these conditions, despite all receiving CCP treatment. These findings align with results from the multivariate model reported by Wu et al.\cite{wu2024bayesian}, which suggested that CCP treatment was more effective for patients with preexisting disease, given the reference levels of other pre-treatment characteristics. However, as shown in Figure 2 in main text, the contribution of these effects was minimal, making their clinical significance uncertain.

\clearpage

\section*{E. Application Results on the Log Scale}

Table \ref{tab2} below presents the log ORs and their corresponding 95\% confidence intervals for each predicted likely responder subgroup, evaluated with respect to four binary outcomes. 
The predicted estimand is the median of point estimate and confidence interval across 100 Monte Carlo (MC) procedures. The reported metrics in Table are as follow: 
1) $\mbox{log OR}$: median log OR; 
2) $\mbox{SE}$: median standard error of log OR; and 
3) $\mbox{95\% CI}$; median 95\% confidence interval of log OR.

\begin{table}[h]
\centering
\begin{threeparttable}
\caption{Each CV-``evaluation" set consists of 1,692 samples. The naïve frequentist method, using XGBoost trained on the CV-``design" set, yields median (SD) sample sizes of 643 (187) for likely-responders, 667.5 (167) for moderate-responders, and 326 (94) for unlikely-responders. In the two-stage model, BART trained on the CV-``design" set produces sizes of 778 (164), 490 (88), and 403.5 (108) in the CV-``evaluation" set, respectively.}\label{tab2}

%\begin{adjustbox}{max width=\textwidth}
\begin{tabular}{|l| ccc | ccc cc|}
\toprule 
& \multicolumn{3}{c|}{XGBoost \ PBS} & \multicolumn{5}{c|}{BART \ PBS} \\
\cmidrule(lr){2-4} \cmidrule(lr){5-9}
& & \multicolumn{2}{c|}{Naive} & & \multicolumn{2}{c}{Naive} & \multicolumn{2}{c|}{Corrected} \\
\cmidrule(lr){3-4}
\cmidrule(lr){6-7}
\cmidrule(lr){8-9}
 & log OR & SE & 95\% CI 
& log OR & SE & 95\% CI  & SE & 95\% CI \\
\midrule 
\multicolumn{9}{|c|}{Binary outcome:  ventilation or worse at day 14} \\
\cmidrule{1-9}
\begin{tabular}{l}
LR 
\end{tabular} & -0.432 & 0.362  & (-1.134, 0.226)
& -0.335 & 0.319 & (-0.977, 0.263)  
& 0.368 & (-1.055, 0.403) \\
\begin{tabular}{l} 
MR 
\end{tabular} & -0.046 & 0.245  & (-0.520, 0.431) 
& -0.115 & 0.282 & (-0.668, 0.451)  
& 0.426 & (-0.952, 0.738) \\
\begin{tabular}{l} 
UR 
\end{tabular} & -0.085 & 0.259  & (-0.568, 0.417) 
& 0.004 & 0.241 & (-0.451, 0.466)  
& 0.290 & (-0.570, 0.562) \\
\begin{tabular}{l} 
All 
\end{tabular} & -0.162 & 0.149  & (-0.453, 0.129) 
& -0.150 & 0.148 & (-0.440, 0.140)  
& 0.148 & (-0.440, 0.140) \\
\midrule
\multicolumn{9}{|c|}{Binary outcome: hospitalization at day 28} \\
\cmidrule{1-9}
\begin{tabular}{l}
LR 
\end{tabular} & -0.364 & 0.316  & (-0.961, 0.230)
& -0.306 & 0.269 & (-0.836, 0.185)  
& 0.316 & (-0.931, 0.293) \\
\begin{tabular}{l} 
MR 
\end{tabular} & -0.098 & 0.205  & (-0.484, 0.289)
& -0.101 & 0.235 & (-0.565, 0.367)  
& 0.356 & (-0.807, 0.621) \\
\begin{tabular}{l} 
UR 
\end{tabular} & -0.084 & 0.248  & (-0.564, 0.399)
& -0.003 & 0.224 & (-0.437, 0.416)  
& 0.274 & (-0.566, 0.505) \\
\begin{tabular}{l} 
All 
\end{tabular} & -0.172 & 0.129  & (-0.425, 0.081)
& -0.158 & 0.128 & (-0.410, 0.094)  
& 0.128 & (-0.410, 0.094) \\
\midrule
\multicolumn{9}{|c|}{Binary outcome: ventilation or worse at day 28} \\
\cmidrule{1-9}
\begin{tabular}{l}
LR 
\end{tabular} & -0.356 & 0.356  & (-1.005, 0.293)  
& -0.313 & 0.311 & (-0.905, 0.309)  
& 0.364 & (-1.010, 0.438) \\
\begin{tabular}{l} 
MR 
\end{tabular} & -0.143 & 0.245  & (-0.621, 0.315)  
& -0.206 & 0.278 & (-0.752, 0.340)  
& 0.420 & (-1.043, 0.612) \\
\begin{tabular}{l} 
UR 
\end{tabular} & -0.008 & 0.255  & (-0.499, 0.466)  
& 0.019 & 0.237 & (-0.444, 0.479)  
& 0.284 & (-0.536, 0.550) \\
\begin{tabular}{l} 
All 
\end{tabular} & -0.175 & 0.146  & (-0.461, 0.113)  
& -0.160 & 0.145 & (-0.445, 0.124)  
& 0.145 & (-0.445, 0.124) \\
\midrule
\multicolumn{9}{|c|}{Binary outcome: mortality at day 28} \\
\cmidrule{1-9}
\begin{tabular}{l}
LR 
\end{tabular} & -0.497 & 0.400  & (-1.239, 0.255)  
& -0.419 & 0.365 & (-1.114, 0.253)  
& 0.409 & (-1.198, 0.369) \\
\begin{tabular}{l} 
MR 
\end{tabular} & -0.276 & 0.291  & (-0.823, 0.278)  
& -0.287 & 0.330 & (-0.926, 0.362)  
& 0.495 & (-1.276, 0.689) \\
\begin{tabular}{l} 
UR 
\end{tabular} & -0.048 & 0.272  & (-0.577, 0.462)  
& -0.078 & 0.254 & (-0.569, 0.429)  
& 0.296 & (-0.657, 0.513) \\
\begin{tabular}{l} 
All 
\end{tabular} & -0.260 & 0.164  & (-0.579, 0.061)  
& -0.251 & 0.164 & (-0.571, 0.070)  
& 0.164 & (-0.571, 0.070) \\
\bottomrule
\end{tabular}
%\end{adjustbox}

\textit{Note.} LR: likely-responders; MR: moderate-responders; UR: unlikely-responders; All: all individuals (unstratified);  
“XGBoost PBS” and “BART PBS”: XGBoost or BART was used to estimate PBS for subgrouping.
\end{threeparttable}
\end{table}

%\pagebreak

\clearpage

\section*{F. Additional Analyses for the BART Model in the COMPILE COVID-19 Clinical Trial}

\subsection*{F1. Assumption Checks: Diagnostic Plots}

\begin{figure}[h]
\centering
\includegraphics[width=13cm]{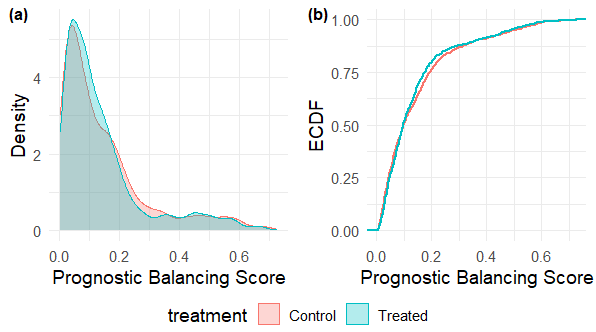}
\caption{Diagnostic comparison of the estimated prognostic score $\hat{s}(x)$ between treatment arms. 
(a) Kernel density estimates show substantial overlap of $\hat{s}(x)$ across treated and control groups. 
(b) Empirical cumulative distribution functions (ECDFs) likewise indicate good overlap. 
These diagnostics support the covariate-overlap assumption underlying the PBS calculation.}
\label{diagnostic}
\end{figure}

To assess whether the prognostic score mapping $x \rightarrow s(x)$ generalizes well across arms, we compared the empirical distributions of $\hat{s}(x)$ for treated and control individuals for COMPILE COVID-19 clinical trial. Figure~\ref{diagnostic} displays density and ECDF plots of the prognostic score by arm, both of which show substantial overlap between groups. 

%\pagebreak

\subsection*{F2. Sensitivity: Sample Size Allocation in Design and Evaluation Subsets}

In the main text, each CV-``design'' set consists of 595 samples (about 600); Each CV-``evaluation’’ set consists of 1,692 samples. 
To assess sensitivity to sample size allocation between the design and evaluation subsets, we re-ran the analysis using smaller ($n=500$) and larger ($n=700$) CV-``design’’ sets. The relative proportions of the LR, MR, and UR subgroups remained stable across these split scenarios (Table~\ref{mccv.split}), and the downstream causal effect estimates within each subgroup were similar (Table~\ref{eff.split}).

\begin{table}[h]
\centering
\caption{Median (SD) of LR, MR, and UR subgroups across 100 MC-CV splits.}\label{mccv.split}
\begin{tabular}{lcccc}
\toprule
CV-``design'' & CV-``evaluation'' & LR & MR & UR \\
$n$ & $n$ & Median (SD) & Median (SD) & Median (SD) \\
\midrule
500 & 1787 & 811 (224)   & 531 (131)  & 435.5 (132.5) \\
595 & 1692 & 778 (164)   & 490 (88)   & 403.5 (108) \\
700 & 1587 & 734.5 (125.5) & 450 (71) & 396 (86) \\
\bottomrule
\end{tabular}
\end{table}

\begin{table}[h]
\centering
\begin{threeparttable}
\caption{Estimated OR and standard errors of log OR for LR, MR, UR, and overall sample across different \textbf{CV-``design’’ set} sizes (n=500, 595, 700). Outcome: ventilation or worse at day 14.}
\label{eff.split}
\begin{tabular}{l r rr rr}
\toprule
 & \multirow{2}{*}{OR} & \multicolumn{2}{c}{Naive} & \multicolumn{2}{c}{Corrected} \\
\cmidrule(lr){3-4}\cmidrule(lr){5-6}
 & & $SE_{logOR}$ & OR 95\% CI  & $SE_{logOR}$ & OR 95\% CI \\
\midrule
\multicolumn{6}{l}{\textbf{n=500 (CV-``design'')}}\\
\midrule
LR  & 0.743 & 0.303 & $(0.407,\, 1.335)$ & 0.385 & $(0.350,\, 1.550)$ \\
MR  & 0.836 & 0.278 & $(0.497,\, 1.432)$ & 0.407 & $(0.381,\, 1.879)$ \\
UR  & 0.977 & 0.227 & $(0.628,\, 1.523)$ & 0.273 & $(0.557,\, 1.689)$ \\
All & 0.856 & 0.141 & $(0.649,\, 1.130)$ & 0.141 & $(0.649,\, 1.130)$ \\
\addlinespace
\midrule
\multicolumn{6}{l}{\textbf{n=595 (CV-``design'')}}\\
\midrule
LR  & 0.715 & 0.319 & $(0.376,\, 1.301)$ & 0.368 & $(0.348,\, 1.496)$ \\
MR  & 0.891 & 0.282 & $(0.513,\, 1.570)$ & 0.426 & $(0.386,\, 2.093)$ \\
UR  & 1.004 & 0.241 & $(0.637,\, 1.593)$ & 0.290 & $(0.566,\, 1.754)$ \\
All & 0.861 & 0.148 & $(0.644,\, 1.150)$ & 0.148 & $(0.644,\, 1.150)$ \\
\addlinespace
\midrule
\multicolumn{6}{l}{\textbf{n=700 (CV-``design'')}}\\
\midrule
LR  & 0.708 & 0.330 & $(0.363,\, 1.358)$ & 0.393 & $(0.323,\, 1.521)$ \\
MR  & 0.906 & 0.305 & $(0.506,\, 1.631)$ & 0.453 & $(0.379,\, 2.222)$ \\
UR  & 0.981 & 0.257 & $(0.596,\, 1.621)$ & 0.310 & $(0.542,\, 1.791)$ \\
All & 0.864 & 0.158 & $(0.633,\, 1.180)$ & 0.158 & $(0.633,\, 1.180)$ \\
\bottomrule
\end{tabular}
\end{threeparttable}
\end{table}

%\clearpage

\subsection*{F3. Model Fit: Posterior Predictive Checks}

\begin{figure}[h]
\centering
\includegraphics[width=10cm]{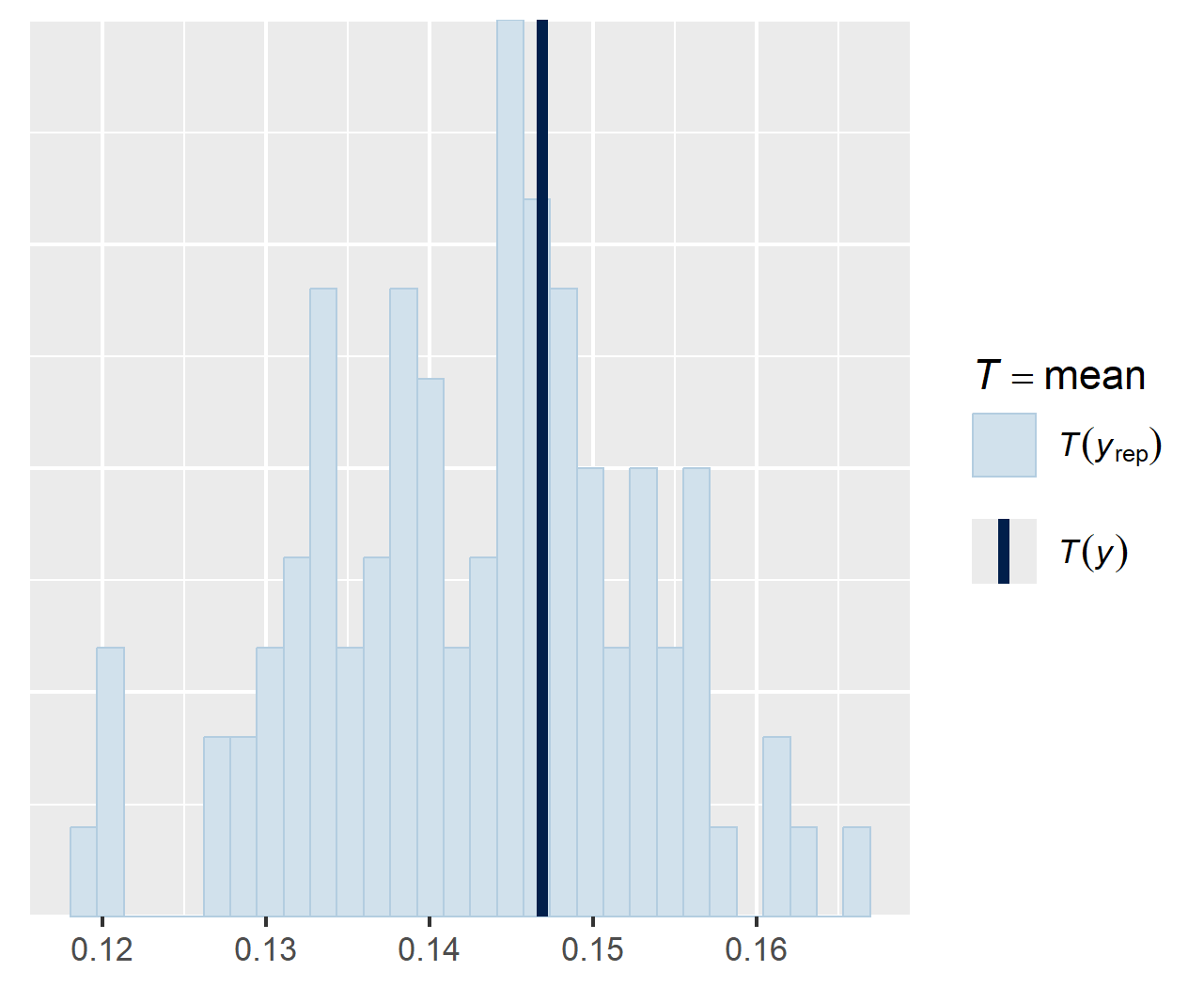}
\caption{Posterior predictive check for treated patients: comparison of observed event rate (vertical line) with posterior predictive distribution of replicated event rates.}
\label{ppc}
\end{figure}

We generated 100 posterior predictive replicates of the treated data. The observed event rate was 0.147, while the posterior predictive mean event rate across replicates was 0.143, with a posterior predictive p-value of 0.38.
Figure~\ref{ppc} displays the posterior predictive distribution compared with the observed data.

%\clearpage

\subsection*{F4. Posterior-Mean Rule for Subgroup Assignment}

\begin{figure}[h]
\centering
\includegraphics[width=10cm]{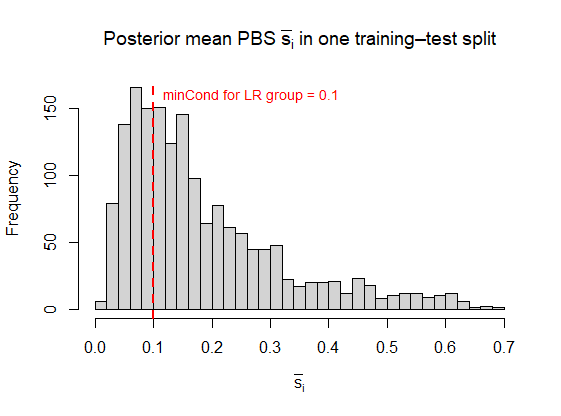}
\caption{Distribution of posterior mean PBS $\bar{s}_i$ across 1,692 subjects in the test subset (mean = 0.183, SD = 0.135).}
\label{mccv_si}
\end{figure}

For implementation and reporting, subgroup membership is assigned using the posterior mean of prognostic-balancing score (PBS), 
\[
\bar{s}_i = \frac{1}{K} \sum_{k=1}^K PBS_i^{(k)},
\]
compared against the prespecified threshold \texttt{minCond}. 
This yields a single rule (“classify by posterior mean”), ensuring that subgroup definitions remain explicit while the second-stage inference continues to propagate the underlying uncertainty. 
Alongside the point classification, we recommend reporting
\[
\pi_i^{\mathrm{LR}} = \frac{1}{K} \sum_{k=1}^K I\!\left(PBS_i^{(k)} \le \texttt{minCond}\right)
\]
as a measure of classification confidence. As an illustration, we applied this rule in the COMPILE analysis.
In one training–test split, the model was trained on $n_{\mathrm{train}} = 595$ (1/4th) subjects and evaluated on the remaining $n_{\mathrm{test}} = 1{,}692$ (3/4th) subjects.
For each subject, the posterior mean PBS, $\bar{s}_i$, was obtained by averaging 100 posterior draws.
Figure~\ref{mccv_si} shows the distribution of $\bar{s}_i$ values (mean = 0.183, SD = 0.135).
With the classification threshold set to $\texttt{minCond} = 0.1$, a total of 539 subjects (out of $1{,}692$ subjects) were classified as likely responders (LR).

\begin{figure}[h]
\centering
\includegraphics[width=15cm]{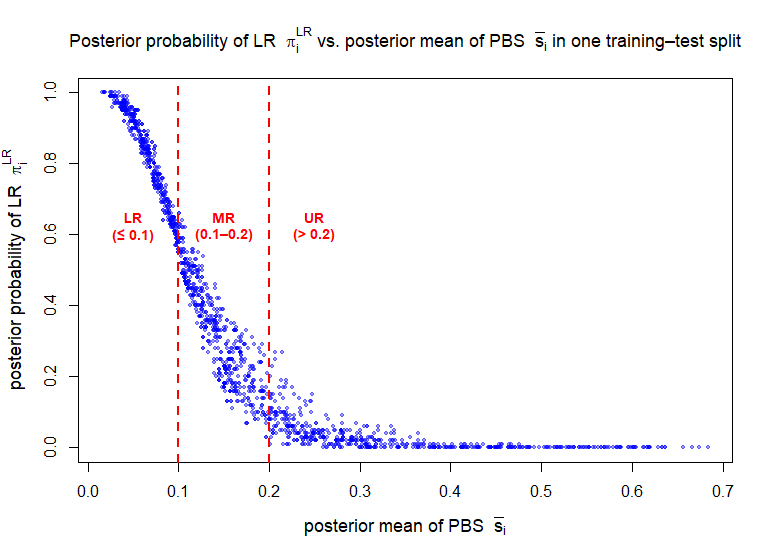}
\caption{Scatter plot of posterior probability of being LR ($\pi_i^{\mathrm{LR}}$) versus posterior mean PBS ($\bar{s}_i$) in one training–test split. Vertical dashed lines indicate subgroup boundaries: $\bar{s}_i \le 0.1$ (LR), $0.1 < \bar{s}_i \le 0.2$ (MR), and $\bar{s}_i > 0.2$ (UR).}
\label{mccv_si_pii}
\end{figure}

We also evaluated the posterior probability of being LR (classification confidence), denoted as $\pi_i^{\mathrm{LR}}$, for each subject in the test set.
Figure~\ref{mccv_si_pii} displays the relationship between the posterior mean PBS ($\bar{s}_i$) and the corresponding LR ``classification confidence''  ($\pi_i^{\mathrm{LR}}$).

Among LR subjects ($n = 539$), the mean classification confidence $\pi_i^{\mathrm{LR}}$  was 0.822 (IQR = 0.220; range: 0.550–1.000).
Among moderate responder (MR) subjects ($n = 583$), the mean was 0.330 (IQR = 0.240; range: 0.030–0.660).
Among unlikely responder (UR) subjects ($n = 570$), the mean was 0.031 (IQR = 0.050; range: 0.000–0.270). High values of $\pi_i^{\mathrm{LR}}$ support confident deployment of LR-based enrichment.

For this LR subgroup (mean $\pi_i^{\mathrm{LR}}\approx 0.82$), the reported subgroup-specific treatment-effect inference is conducted via posterior–design averaging (Stage 2 applied across posterior-induced subgrouping). Consequently, the confidence intervals account for both outcome-model uncertainty and uncertainty in subgrouping.

\clearpage

\section*{G. Application Results for Relative Treatment Effect Subgroup}

We used the primary outcome (the binary indicator for ventilation or worse on day 14) as the outcome in both the ``design'' stage and the ``evaluation'' stage. We implemented a Monte Carlo (MC) cross-validation procedure similar to the analysis for likely responder in main text.
In each MC iteration, we followed the following steps:

\begin{itemize} 
\item Individuals with $T = 1$ and $T = 0$ were each randomly split into two equal-sized subsets. One subset from each group was assigned to the CV-“design” set, while the remaining subsets from both groups formed the CV-“evaluation” set. 
\item We train Bayesian logistic regression model on the CV-``design'' set using the primary outcome, based on pre-treatment characteristics and their interaction with treatment as predictors. The trained model was then applied to individuals in the CV-``evaluation'' set to calculate the OR, where we set the confounder variables at RCT ID = AA and enrollment quarter = Apr–Jun 2020 to standardize confounding effects in the OR calculation. 

\item After categorizing each individual in the CV-``evaluation'' set into subgroups, the OR for subgroup-specific treatment effects was estimated for each subgroup on the primary outcome.
\end{itemize}

A final predicted estimand was obtained by taking the median of point estimate and confidence interval across 100 MC procedures. The reported metrics in Table are as follow: 
1) $\mbox{OR}$: exponentiated median of the log ORs; 
2) $\mbox{SE}_{logOR}$: median of the standard errors of the log ORs; and 
3)  $\mbox{95\% CI}$: exponentiated median of the lower and upper bounds of the 95\% confidence intervals of the log ORs.

In this illustrative analysis, the subgroup stratification was based on the predicted OR of ventilation or worse on day 14 for each individual in the CV-``evaluation'' set. 
We set the treatment effect (CCP vs. control) thresholds at $OR = 0.9$, categorizing individuals in the CV-``evaluation'' set into two groups: 
if \( OR \leq 0.9 \), then "likely to benefit" (LB); 
if \( PBS > 0.9 \), then "unlikely to benefit" (UB). 
Table \ref{tab3} presents the ORs and their corresponding 95\% confidence intervals for each likely to benefit subgroup.

\begin{table}[h]
\centering
\begin{threeparttable}
\caption{Each CV-``evaluation'' set consists of 1,144 samples per replication. The median (SD) sample size is 522 (228) for the likely-benefit group and 622 (228) for the unlikely-benefit group.}\label{tab3}

\begin{tabular}{|l| ccc cc|}
\toprule 
& \multicolumn{5}{c|}{Bayesian Logistic Regression} \\
 \cmidrule(lr){3-6}
 & & \multicolumn{2}{c}{Naive} & \multicolumn{2}{c|}{Corrected} \\
\cmidrule(lr){3-4}
\cmidrule(lr){5-6}
& OR & $SE_{logOR}$ & OR 95\% CI  & $SE_{logOR}$ & OR 95\% CI \\
\midrule 
\multicolumn{6}{|c|}{Binary outcome: the need for ventilation or worse at day 14} \\
\cmidrule{1-6}
\begin{tabular}{l}
UB
\end{tabular} 
& 0.880 & 0.237 & (0.555, 1.407)  
& 0.532 & (0.321, 2.308) \\
\begin{tabular}{l} 
LB
\end{tabular} 
& 0.850 & 0.260 & (0.501, 1.439) 
& 0.416 & (0.377, 1.874) \\
\begin{tabular}{l} 
All 
\end{tabular} 
& 0.862 & 0.171 & (0.616, 1.204)  
& 0.171 & (0.616, 1.204) \\

\bottomrule
\end{tabular}

\begin{tablenotes}[flushleft]
\item Abbreviations: 
\item LB, likely benefit from CCP vs. control; UB, unlikely benefit from CCP vs. control; All, all individuals.
\item $ SE_{logOR}$, the standard error (SE) of the log odds ratio (OR).
\end{tablenotes}
\end{threeparttable}
\end{table}

Using our proposed ``corrected'' method, the confidence intervals are wider compared to the ``naive'' method, reflecting greater uncertainty about the effect of CCP within the data-driven identified subgroups.

\clearpage

%Bibliography
\bibliographystyle{unsrt}  
\bibliography{references}

\end{document}